\documentclass[lettersize,journal]{IEEEtran}
\usepackage{amsmath,amsfonts}
\usepackage{algorithmic}
\usepackage{algorithm}
\usepackage{array}
\usepackage[caption=false,font=scriptsize,labelfont=sf,textfont=sf]{subfig}
\usepackage{textcomp}
\usepackage{stfloats}
\usepackage{url}
\usepackage{verbatim}
\usepackage{graphicx}
\usepackage{cite}
\graphicspath{{figs}}
\usepackage[nameinlink, capitalize]{cleveref}
\usepackage{siunitx}

\usepackage[commandnameprefix=ifneeded]{changes}
\setaddedmarkup{\textcolor{blue}{\uwave{#1}}}
\setdeletedmarkup{\textcolor{red}{\sout{#1}}}

\begin{document}

\title{Effects of Social Contextual Variation Using Partner Avatars on Memory Acquisition and Retention}

\author{Takato Mizuho, Takuji Narumi, and Hideaki Kuzuoka
\thanks{T. Mizuho, T. Narumi, and H. Kuzuoka are with the Graduate School of Information Science and Technology, the University of Tokyo, Tokyo, Japan.}
\thanks{Manuscript received April xx, 20xx; revised August xx, 20xx.}}

\markboth{Journal of \LaTeX\ Class Files,~Vol.~xx, No.~x, August~20xx}%
{Mizuho \MakeLowercase{\textit{et al.}}: A Sample Article Using IEEEtran.cls for IEEE Journals}

\IEEEpubid{0000--0000/00\$00.00~\copyright~2021 IEEE}

\maketitle

\begin{abstract}
This study investigates how partner avatar design affects learning and memory when an avatar serves as a lecturer. Based on earlier research on the environmental context dependency of memory, we hypothesize that the use of diverse partner avatars results in a slower learning rate but better memory retention than that of a constant partner avatar. Accordingly, participants were tasked with memorizing Tagalog--Japanese word pairs. On the first day of the experiment, they repeatedly learned the pairs over six sessions from a partner avatar in an immersive virtual environment. One week later, on the second day of the experiment, they underwent a recall test in a real environment. We employed a between-participants design to compare the following conditions: the varied avatar condition, in which each repetition used a different avatar, and the constant avatar condition, in which the same avatar was used throughout the experiment.
Results showed that participants in the varied avatar condition recalled significantly worse during the learning trials on the first day. However, we found no significant difference between conditions in the delayed recall test on the second day. We discuss these effects in relation to the social presence of the partner avatar.
This study opens up a novel approach to optimizing the effectiveness of instructor avatars in immersive virtual environments. 
\end{abstract}


\section{Introduction}
\IEEEPARstart{W}{hen} interacting with embodied virtual agents and partner avatars in virtual environments, their characteristics influence users' perception and cognition~\cite{Lin-etal2023, Volonte-etal2016, Zibrek-etal2019}.
An embodied virtual agent is typically defined as a social entity controlled by a computer program or algorithm, whereas an avatar encompasses a broader definition, including entities controlled by recorded or real-time human actions. In this paper, we adopt the term partner avatar, as the term avatar in the field of virtual reality (VR) is often used in a broader sense to refer to various types of entities within virtual environments.
Especially, previous studies have shown that a partner avatar's appearance and behavior can affect the user's memory in learning and training scenarios where the avatar acts as an instructor~\cite{Amemiya-etal2022, Mizuho-etal2023ahs, Petersen-etal2021}. These effects occur because the agent or avatar establishes a social relationship with the user rather than being a mere image on a screen~\cite{Reeves-Nass1996}. A key objective of the research conducted in this field has been to identify the avatar designs that optimize learning and training outcomes.

In line with such research trends, this study also investigates the effect of the appearance of partner avatars on memory. The novelty of our study is that it analyzes the effect from the perspective of the environmental context-dependency of memory.
Environmental context dependency refers to the property whereby memory is influenced by incidental environmental contexts such as location and the presence of other people~\cite{Smith2013, Smith-Vela2001}. Representative effects include the reinstatement effect, which promotes recall when the learning and recall contexts are the same, and the multiple-context effect, which facilitates memory retention when learning occurs across diverse contexts compared to a single context. Due to the environmental context dependency of memory, transitioning from a virtual to a real-world environment causes forgetting~\cite{Lamers-Lanen2021}. In other words, materials learned in VR may not be easily recalled in the real world. Therefore, understanding how the environmental context of VR influences memory is critical for improving the effectiveness of VR-based educational applications.

\IEEEpubidadjcol

Previous studies demonstrated that using VR to manipulate the surrounding environment~\cite{Mizuho-etal2024tvcg1, Shin-etal2021} or a virtual body (self-avatar)~\cite{Mizuho-etal2024tvcg2} can induce the reinstatement and multiple-context effects as in the physical world. Extending these findings, this study investigates the effects of partner avatars in an immersive virtual environment (IVE). We posited that altering the partner avatar's appearance would change the social context of who you are with, thereby eliciting context-dependent memory effects. While prior research in the physical world has acknowledged the influence of social context, researchers have largely underexplored this topic due to the difficulties in fully controlling the appearance and behavior of other people. In this regard, VR technology enables us to overcome these limitations by enabling precise control of such factors, thereby facilitating rigorous investigations. Motivated by this idea, earlier studies examined the effect of screen-based avatars on memory~\cite{Mizuho-etal2023ahs, Mizuho-etal2024sap}; however, their findings were inconclusive. Hence, in this study, we investigate the effect of partner avatars in an IVE using a head-mounted display (HMD) instead of on-screen avatars. We expected that the increased immersion provided by the HMD would enhance the social presence of the partner avatars~\cite{Guimaraes-etal2020}, thereby amplifying their effect on memory to produce more observable results. This research represents the first investigation into the role of partner avatars as environmental context affecting memory within IVEs.

\begin{figure*}[tb]
 \centering
 \includegraphics[width=\linewidth]{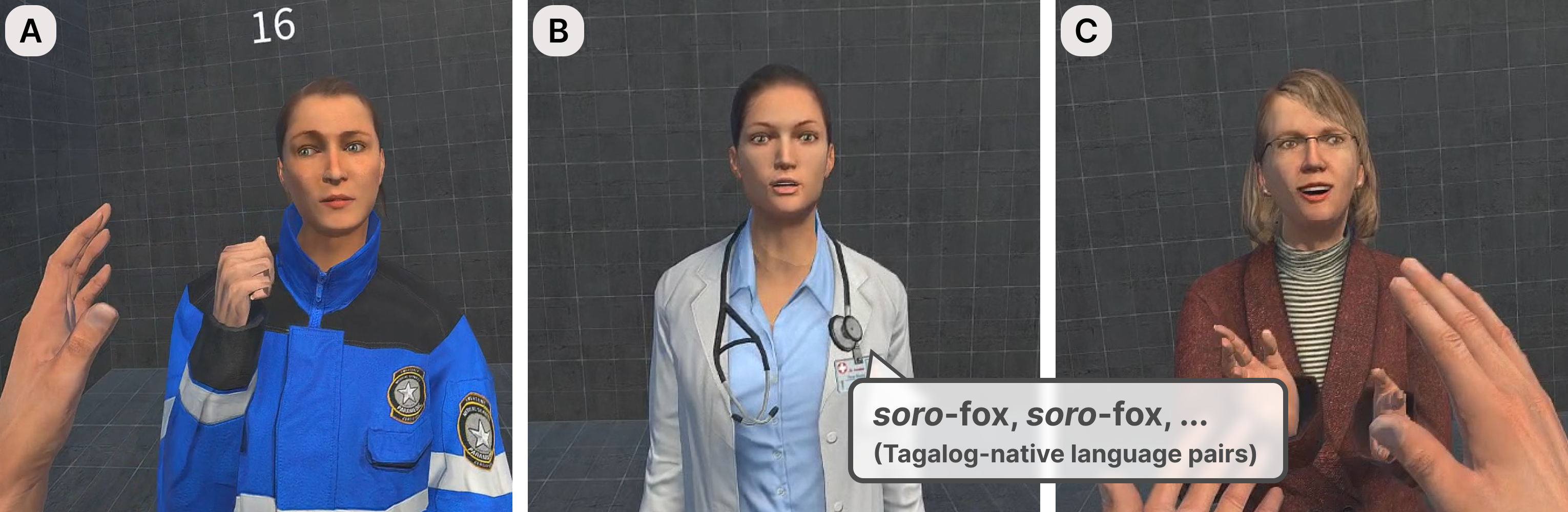}
 \caption{Participants learned Tagalog-Japanese word pairs from a lecturer avatar within an immersive virtual environment. These images depict first-person perspective views of the participant, showing the partner avatar's actions and facial expressions during each phase of a learning trial: (A) during the avatar familiarization phase, (B) during word presentation and recall tests, and (C) at the conclusion of the trial. Note that the avatar remained constant within each trial and could change between trials.}
 \label{fig:teaser}
\end{figure*}

\begin{figure*}[tb]
 \centering
 \includegraphics[width=\linewidth]{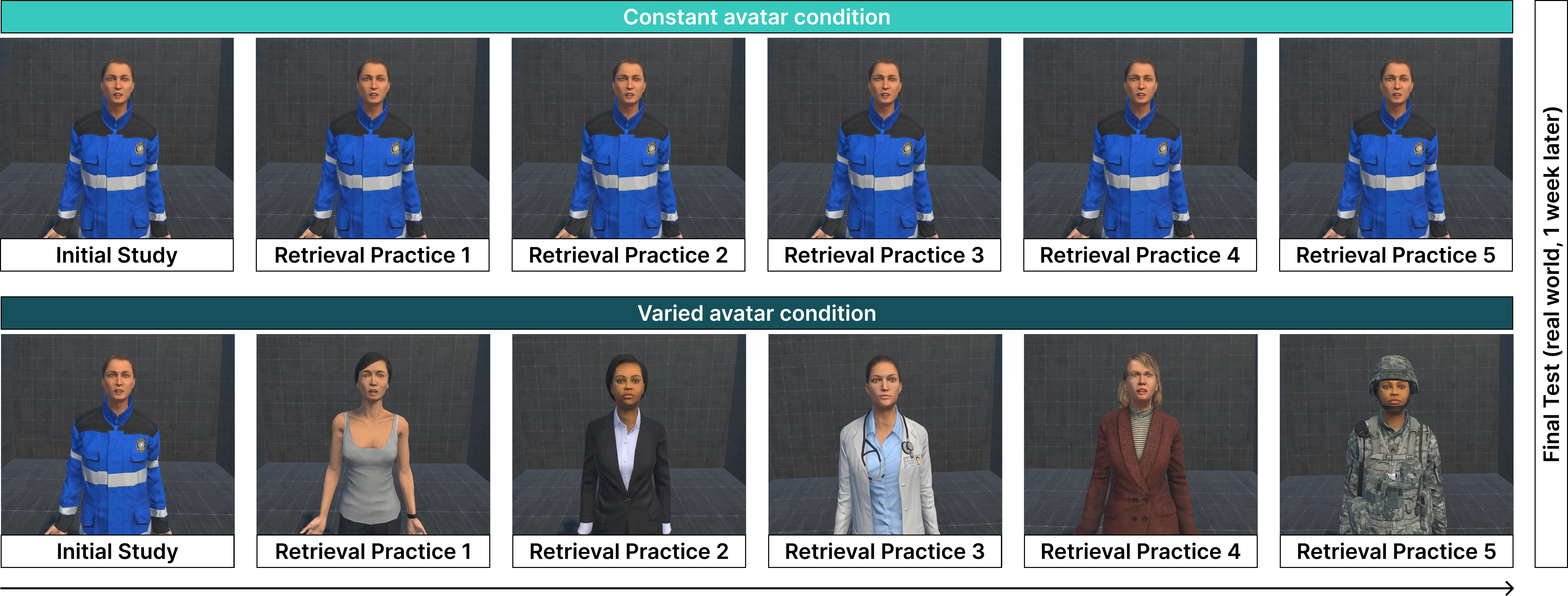}
 \caption{Participants performed an initial study trial, in which they learned 20 Tagalog--Japanese pairs spoken by the avatar, and five retrieval practice (RP) trials, in which they were tested and received feedback. We compared the learning curves during RP1--5 and memory retention during a final test conducted in a real environment one week later between two conditions: the constant avatar condition, where the same partner avatar was used for all six trials, and the varied avatar condition, where a different partner avatar was used for each trial.}
 \label{fig:design}
\end{figure*}

In this study, participants performed a memory task involving remembering the Tagalog--Japanese pairs spoken by a partner avatar (\Cref{fig:teaser}). On the first day, the participants learned the word pairs during an initial study (IS) trial. Subsequently, they advanced their learning in five retrieval practice (RP1--5) trials, during which they attended tests and received feedback. One week later, they took the final test (FT) in a real environment. The experiment adopted the following two-condition, between-participants design (\Cref{fig:design}): a constant avatar condition, in which the same avatar was used throughout the IS and RP1--5 trials, and a varied avatar condition, in which different avatars were used for each trial. We hypothesized that recall performance during RP1--5 would be higher under the constant condition than the varied one due to the reinstatement effect. Conversely, we hypothesized that recall performance in the FT would be higher under the varied condition than the constant one, reflecting the multiple-context effect. Our results showed the partner avatar's significant reinstatement effect but did not provide any evidence of a multiple-context effect.

The contributions of our study are as follows: (1) We demonstrated for the first time that partner avatars in an IVE can affect memory as an environmental context. Specifically, we showed a significant reinstatement effect of the partner avatars, according to which changing the instructor avatar's appearance during repeated learning inhibits memory acquisition. (2) However, we could not observe a multiple-context effect of the partner avatars, suggesting that the social contextual variation during learning may have little impact on memory retention one week later in the physical world.
These findings, along with earlier results related to IVEs and self-avatars, contribute to a deeper understanding of how environmental context in VR affects memory and help expand its potential applications. Additionally, (3) correlation analyses suggested the involvement of the social presence of partner avatars in these effects. The participants who perceived a higher social presence might have been more likely to experience the context-dependent memory effects. This trend suggests that enhancing social presence may be a promising approach to achieving further memory enhancement.

\section{Related Work}
\subsection{Embodied Agents and Partner Avatars}
A social entity controlled by a computer is typically referred to as an agent, while one controlled by a human is known as an avatar~\cite{Bailenson-Blascovich2004, Weidner-etal2023}. In the field of VR, however, the term avatar is often broadly defined to refer to diverse entities within virtual environments, regardless of whether they are computer-controlled or human-controlled~\cite{Yuan-Gao2024, Lin-etal2023}. In this paper, the virtual humans used in the experiment were computer-controlled to ensure high reproducibility.
Accordingly, we adopt the term avatar in this study to refer to the visual representation of these entities, aligning with its broader usage. However, we acknowledge that the perception of agency (i.e., whether an entity is controlled by a human or a computer) is a critical factor in social interaction~\cite{Kyrlitsias-MichaelGrigoriou2022}, which we address in the Limitations section.

Researchers propose the implementation of remote or asynchronous lectures using video materials, in which embodied virtual agents or avatars act as instructors. They have examined the effects of an avatar's realism~\cite{Baylor-Kim2004, Horovitz-Mayer2021, Hsieh-Sato2021}, anthropomorphism (e.g., human vs. robot or animal)~\cite{Li-etal2016, Mizuho-etal2024sap}, user preference~\cite{Amemiya-etal2022, Pi-etal2022}, and congruence with content~\cite{Beege-etal2022} on learning outcomes. These effects are attributed to the agent's ability to establish human-like social relationships with users~\cite{Reeves-Nass1996}, rather than being perceived as mere images.


While most earlier studies have focused on the effects of virtual agents presented on 2D monitors, extending these findings to IVEs is also a promising area of research. Immersive experiences using HMDs have been suggested to enhance the social presence of virtual agents~\cite{Guimaraes-etal2020}, potentially amplifying their influence on users. However, this field remains in its early stages and lacks a comprehensive understanding. Petersen et al.~\cite{Petersen-etal2021} suggested that the use of virtual humans can distract attention from learning materials, resulting in poorer memory performance for factual knowledge than an audio-only condition. They also examined the role of realism in the appearance and behavior of virtual humans, suggesting that while behavioral realism significantly impacts social presence, it has minimal influence on memory performance. Rzayev et al.~\cite{Rzayev-etal2019} compared various conditions, including humans in real environments, virtual humans in VR, robots in VR, and audio-only in VR, indicating that understanding of content delivered by avatars did not significantly differ across these conditions. Makransky et al.~\cite{Makransky-etal2019JCAL} conducted a study with junior high school students, demonstrating that a drone-type agent and a young female researcher agent improved learning outcomes for male and female students, respectively, highlighting the potential benefits of personalization.

On the other hand, previous studies have explored the effects of partner avatars in IVEs on users' perceptions, which could potentially influence memory performance.
These studies examined the effects of appearance features, such as rendering style, photorealism, and visibility, on social presence, trustworthiness, and uncanny valley effect~\cite{Aseeri-Interrante2021, Lin-etal2023, Zibrek-etal2018, Zibrek-etal2019}. Additionally, they explored how personality traits expressed through body movements and facial expressions~\cite{Patotskaya-etal2023, Zibrek-etal2018, Zibrek-etal2020} and voice~\cite{Higgins-etal2022} affect user perceptions.
Whereas most studies used questionnaires to assess subjective evaluations, some used behavioral measures. A commonly used measure is proximity, which is defined as the shortest distance between the avatar and the user during a task. This measure is based on human beings' tendency to maintain a comfortable distance from others, which varies depending on whether interactions involve virtual objects or virtual humans~\cite{Sanz-etal2015}. Therefore, proximity can reveal whether a user perceives the avatar as a human-like entity or merely a virtual object.

In this study, we investigated the effective partner avatar designs supporting learning and memory in VR-based education and training. Specifically, we introduced a new approach to examine how appearance changes during repeated learning affect memory based on its environmental context dependency (\Cref{sec:context-dependent-memory}). Additionally, drawing on previous research on IVE avatars, we measured social presence and proximity to know how users perceive avatars and explored the potential relationship between these indices and memory performance. Participants were tasked with memorizing words spoken by partner avatars. Based on the study by Kruse et al.~\cite{Kruse-etal2023SUI}, we framed this task as one requiring active engagement with the avatar, ensuring participants interacted with the avatar rather than ignoring it during the task.

\subsection{Environmental Context Dependency of Memory}
\label{sec:context-dependent-memory}

The memories of subjective experiences are called episodic memories~\cite{Tulving-Thomson1973}, of which environmental context dependency is an important characteristic~\cite{Smith-Vela2001}. Environmental context dependency is the property whereby incidental environmental contexts, such as place, background music, or odors, influence memory encoding, retention, and retrieval. The reinstatement and the multiple-context effects are two well-known effects.
The reinstatement effect is an effect in which recall performance is better when the environmental contexts during learning and testing are identical than when they differ. For example, in the well-known study by Godden and Baddeley~\cite{Godden-Baddeley1975}, the participants who learned a word list underwater recalled the words more effectively on being tested underwater than on land. Similarly, the participants who learned the words on land recalled them better on land than underwater.

The multiple-context effect is an effect of repeated learning under diverse environmental contexts that makes memory resistant to contextual changes and reduces forgetting. For example, Smith et al.~\cite{Smith-etal1978} compared a varied room condition, under which a word list was learned once in each of two different rooms, and a constant room condition, under which the list was learned twice in the same room. They found that, compared to the constant condition, the varied condition exhibited significantly higher recall scores on a test conducted in a new room. Further, Smith and Handy~\cite{Smith-Handy2014, Smith-Handy2016} conducted experiments using background photographs and videos on a monitor. They compared a varied context condition, where a different background was used for each repetition, and a constant context condition, where the same background was used consistently during the repeated learning of face--name or Tagalog--English pairs. Results showed that memory acquisition during repeated learning was significantly higher in the constant condition than the varied condition due to the reinstatement effect. However, after an interval (5 minutes or 2 days), the varied condition demonstrated less forgetting and higher memory retention during the final test.

The mechanism underlying the multiple-context effect is not yet completely clarified; however, research proposes two hypotheses. The first is the encoding variability hypothesis~\cite{Bower1972}, which suggests that in a varied context condition, different contextual information is associated with the memory of the words in each repetition. This increases the number of contextual cues available at retrieval and, thereby, enhances recall. The second hypothesis is the desirable difficulty hypothesis~\cite{Bjork-Bjork1992},
according to which a large cognitive load for retrieval in a varied context condition due to the reinstatement effect can strengthen memory trace, making later recall easier.

This study manipulated partner avatars as environmental contexts to examine the reinstatement and multiple-context effects. We posited that changing the partner avatar would alter the social context. Although previous research in real environments indicates that social context influences memory~\cite{Isarida-Isarida2010}, direct verification is challenging due to the difficulty in completely controlling the appearance and behavior of others. By using VR, this study rigorously controlled various confounding factors associated with partners, thereby pioneering research in an area that was previously difficult to investigate. As mentioned in \cref{sec:vr}, VR's ability to freely edit the surrounding environmental context enhances our understanding of memory characteristics.

\subsection{Reinstatement and Multiple-Context Effects Using Virtual Reality}
\label{sec:vr}
Lamers and Lanen~\cite{Lamers-Lanen2021} identified an important limitation of VR-based learning from the perspective of the environmental context dependency of memory. According to them, switching between virtual and real environments causes context-dependent forgetting due to the reinstatement effect. Specifically, the information learned in VR was easily recalled within VR but difficult to retrieve in the real world, and vice versa. Therefore, to maximize the effectiveness of VR-based education, it is essential to understand the environmental context dependency of memory within VR.

Shin et al.~\cite{Shin-etal2021} were the first to demonstrate the reinstatement effect in IVEs using Martian and underwater IVEs. Their findings suggested that an audiovisual IVE induced an environmental context-dependent effect on memory, even though the physical context surrounding the participants did not change. However, previous studies have yielded mixed results in this respect: whereas some found a significant reinstatement effect~\cite{Koch-Coutanche2024}, others did not~\cite{Chocholackova-etal2023, DeBack-etal2018, Mizuho-etal2023tvcg, Mizuho-etal2023vrst, Walti-etal2019}, indicating that the issue requires further investigation.
Further, the multiple-context effect in IVEs was first demonstrated by Mizuho et al.~\cite{Mizuho-etal2024tvcg1} using 360-degree videos. They showed that repeated learning of Tagalog--Japanese pairs under diverse IVEs significantly reduced forgetting on the final test conducted two days later compared to the constant IVE condition.

Whereas most of the earlier VR-based studies focused on the surrounding scenery, Mizuho et al.~\cite{Mizuho-etal2024tvcg2} investigated the impact of the self-avatar on memory. In memorizing sign-language gestures and corresponding words, they compared a varied avatar condition, where participants used different self-avatars for repeated learning, and a constant avatar condition, where the participants used a single self-avatar throughout. Results showed that the varied avatar condition temporarily reduced memory acquisition due to the reinstatement effect. However, they showed significantly higher memory retention in the final test conducted in a real environment one week later, which was attributable to the multiple-context effect.

In this study, we examined the reinstatement and multiple-context effects of partner avatars by extending previous research on IVEs and self-avatars. We hypothesized that manipulating the appearance of the partner avatar would change the social aspect of the environmental context (social context), eliciting these context-dependent effects in memory. We adopted the experimental design proposed by Mizuho et al.~\cite{Mizuho-etal2024tvcg1, Mizuho-etal2024tvcg2}. While several studies used screen-based avatars~\cite{Mizuho-etal2023ahs, Mizuho-etal2024sap} or scenery combined with other avatars~\cite{Takenaka-etal2024}, this study was the first to investigate the pure effects of partner avatars in IVEs. As IVEs, self-avatars, and other avatars are key components of VR experiences, a clear understanding of their effects on memory can broaden the range of applicable VR experiences. Hence, this study aims to advance the current understanding of the environmental context dependency of memory in VR and to provide insights to improve the effectiveness of VR-based learning.

\section{Experiment}
\subsection{Design}
We designed our experiment based on the previous studies by Smith and Handy~\cite{Smith-Handy2014, Smith-Handy2016} and Mizuho et al.~\cite{Mizuho-etal2024tvcg1, Mizuho-etal2024tvcg2}.
On the first day of the experiment (Day 1), participants participated in an IS trial in which they memorized the Tagalog--Japanese word pairs. Subsequently, they completed five RP trials (RP1--5), during which they were tested and received feedback. One week later, on the second day (Day 2), the participants underwent an FT trial in a real environment, which assessed their retention and forgetting.

We compared constant and varied avatar conditions using a between-participants design (\Cref{fig:design}). In the constant avatar condition, the same avatar was used throughout the six VR trials (IS and RP1--5). In contrast, the varied avatar condition involved using a different avatar for each trial. We primarily compared the proportion of correct answers in RP1--5 and FT.

\subsection{Participants}
Twenty-four participants (12 females and 12 males, aged 24.25 $\pm$ 2.75 (\textit{SD}) years) took part in the experiment. They were recruited via social networks and included both students and working adults. All had normal or corrected-to-normal vision. To control for familiarity with memory targets (\Cref{sec:words}), we recruited native Japanese speakers alone. Additionally, the participants clarified their familiarity with VR using a four-point questionnaire with the following responses: (1) none, (2) a few times, (3) a dozen times, and (4) more times. The average score was 2.42 (\textit{SD} = 1.02). Specifically, 3 participants had no prior experience, 14 had used VR a few times, 1 had experienced it a dozen times, and 6 had extensive VR experience.
The participants received compensation with an Amazon gift card amounting to approximately \$20 USD. The experimental protocol was approved by the local ethics committee of the Graduate School of Information Science and Technology, the University of Tokyo (UT-IST-RE-230616).

We determined the sample size as follows. The overall sample size had to be a multiple of 12 to ensure that the avatar types used in the constant condition were counterbalanced across participants.
Using G*Power~\cite{Faul-etal2007}, with the primary measure defined as the difference in the amount of forgetting from Day 1 to Day 2 between the two groups, we found that the sensitivity (\textit{t} tests, \textit{Means: Difference between two independent means}, two-tailed test, $\alpha$ = 0.05, power ($1 - \beta$) = 0.80) was \textit{d} = 1.79 for \textit{N} = 12, \textit{d} = 1.20 for \textit{N} = 24, and \textit{d} = 0.96 for \textit{N} = 36.
Further, we reviewed the effect sizes of the multiple-context effect reported in previous studies: \textit{d} = 1.57, 2.16, and 2.13 in the experiments by Smith and Handy~\cite{Smith-Handy2014}, who used the background 2D videos; \textit{d} = 0.81 in the study by Mizuho et al.~\cite{Mizuho-etal2024tvcg1} that employed 360-degree video-based IVEs; and \textit{d} = 1.49 in the study by Mizuho et al.~\cite{Mizuho-etal2024tvcg2} that used self-avatars. Based on these effect sizes, we expected a large effect size and determined an appropriate sample size of \textit{N} = 24. This sample size was consistent with the previous study using the same experimental design~\cite{Mizuho-etal2024tvcg2}. We also reported the statistical power based on the effect sizes because the observed effect sizes could differ from our expectations.

We manually assigned participants to each condition in a pseudo-random manner, with group-level homogeneity in gender, age, and prior VR experience taken into account.
Consequently, the constant condition included six males and six females, having an average age of 23.8 (\textit{SD} = 1.47) years and an average VR familiarity score of 2.33 (\textit{SD} = 1.07). In contrast, the varied condition included six males and six females, with an average age of 24.7 (\textit{SD} = 3.65) years and an average VR familiarity score of 2.5 (\textit{SD} = 1.00).

\subsection{System}
\subsubsection{Apparatus}
On Day 1, participants experienced VR using a Windows-based PC (AMD Ryzen 7 6800H with Radeon Graphics, 16GB RAM, and NVIDIA GeForce RTX 3070 Ti Laptop GPU), a Meta Quest 3 HMD in Quest Link mode (target frame rate: 90 fps), and Sony WH-1000XM3 noise-canceling headphones. The virtual environment was rendered using Unity 2021.3.26f1 and displayed through the HMD.
Participants were embodied with self-avatars. We chose a male avatar (Male\_Adult\_08) for male participants and a female one (Female\_Adult\_01) for female participants from the Microsoft Rocketbox Avatar Library~\cite{GonzalezFranco-etal2020frontiers}. These avatars were selected for their neutrality and consistency with previous studies~\cite{Mizuho-etal2024tvcg1}. They also differed from the partner avatars used in the study (\Cref{sec:other-avatar}). Participants' head and hand movements were tracked using the HMD's built-in cameras and reflected on the self-avatars, utilizing inverse kinematics calculations using the RootMotion FinalIK plug-in\footnote{\url{https://assetstore.unity.com/packages/tools/animation/final-ik-14290}} and the Microsoft MoveBox-for-MicrosoftRocketbox framework~\cite{GonzalezFranco-etal2020aivr}.
Throughout the experiment, the HMD recorded participants' first-person perspective views and microphone input. We reviewed these recordings to score the participants' oral responses in the memory tests.

On Day 2, the participants performed a memory test in the physical world. They were seated in front of a monitor and required to respond with the Japanese words corresponding to the Tagalog words presented aloud through a Bose Soundlink Revolve II speaker. The monitor displayed a countdown starting from 5 seconds for each word to indicate the time limit of the response. We recorded the participants' responses during the test using an external camera and reviewed the recordings to score them.

\subsubsection{Partner Avatars}
\label{sec:other-avatar}
We selected six female avatars (\Cref{fig:design}) from the Microsoft Rocketbox Avatar Library~\cite{GonzalezFranco-etal2020frontiers}. We used this library because it was open-source and ensured reproducibility. Moreover, it was used by a previous study on self-avatars~\cite{Mizuho-etal2024tvcg2}. The avatars were chosen to be distinct from each other in meaning and clothing color to change the environmental context and effectively induce the desired effects. This selection was similar to the one performed in the previous study~\cite{Mizuho-etal2024tvcg2}. The six selected avatars were a paramedic dressed in blue (Medical\_Female\_03), an instructor dressed in sportswear (Sports\_Female\_02), a businesswoman in a black suit (Business\_Female\_01), a medical doctor in a white coat (Medical\_Female\_02), an elderly person dressed in a red suit (Business\_Female\_02), and a military person in a green camouflage uniform (Military\_Female\_02).
The use of female avatars alone was a major limitation of this study.
However, to our knowledge, previous studies on partner avatars often used avatars of one gender~\cite{Lin-etal2023, Patotskaya-etal2023, Zibrek-etal2018, Zibrek-etal2019}. Therefore, we decided to conduct an experiment using female avatars alone and discuss the possible impact of the avatar's gender on the outcomes.

To ensure reproducibility, the partner avatar's movements were implemented using animation clips from the Microsoft Rocketbox Avatar Library~\cite{GonzalezFranco-etal2020frontiers}. These clips included an idling motion (f\_idle\_neutral\_01), a greeting motion at the beginning of the experiment (f\_gestic\_talk\_self-assured\_01), a waiting motion during proximity measurements (f\_gestic\_thoughtful\_01), a speaking motion during word presentation (f\_gestic\_listen\_accept\_02), and a greeting motion at the end of the experiment (f\_claphands\_01). These clips, which were intended for female characters, mainly featured upper body movements and did not involve movement from the spot; however, they included facial expressions and finger movements. They were selected through informal pilot testing to match the experimental situations.
Consistent with previous studies~\cite{Lin-etal2023, Zibrek-etal2019}, the head direction was adjusted to ensure the avatar looked at and followed the participant, which enhances the perception of social presence. Additionally, the avatar's mouth was lip-synced to the audio file (\Cref{sec:words}) using the Oculus Lipsync plug-in\footnote{\url{https://developer.oculus.com/downloads/package/oculus-lipsync-unity/}}.

In the constant avatar condition, the type of avatar used was balanced across participants by assigning each character to two participants. Conversely, in the varied avatar condition, the order of avatar presentation was randomized individually for each participant using a unique permutation of the six avatars. Although this approach did not constitute a fully counterbalanced design, it reduced systematic bias in avatar order.

\subsubsection{Words}
\label{sec:words}
Participants learned 20 Tagalog--Japanese pairs.
Tagalog is a major language spoken in the Philippines and forms the basis of Filipino, the country's standardized national language.
Tagalog was selected because it uses the Latin alphabet and has a recognizable pronunciation system, similar to English, making its spoken forms relatively easy for Japanese participants to process. Additionally, participants had limited prior knowledge of the language. Previous studies~\cite{Mizuho-etal2024tvcg1, Smith-Handy2016} also used Tagalog words.
We selected 20 word pairs as follows: First, we extracted two-syllable Tagalog nouns from a Filipino\footnote{Filipino is an official language and a standardized version of Tagalog.}--Japanese dictionary. We included only those words whose corresponding Japanese nouns were two to four letters long. We then excluded the words that could be easily inferred from Japanese or English. Additionally, we balanced the initial letters to ensure that they were not too easily distinguishable. The final selection included \textit{ahas} (snake), \textit{apo} (grandchild), \textit{baga} (lung), \textit{bilog} (circle), \textit{dagat} (sea), \textit{dugo} (blood), \textit{hari} (king), \textit{himig} (song), \textit{itim} (black), \textit{init} (heat), \textit{kahon} (box), \textit{kotse} (car), \textit{lamok} (mosquito), \textit{lobo} (wolf), \textit{pisi} (string), \textit{pating} (shark), \textit{sino} (who), \textit{soro} (fox), \textit{timog} (south), and \textit{tulay} (bridge).

Both Tagalog and Japanese were presented audibly through the partner avatar. We chose this presentation method because an informal pilot test suggested that the presentation of Tagalog in audio and Japanese in text increased cognitive load and distracted attention from the avatar. We used Ondoku\footnote{\url{https://ondoku3.com/}} text-to-speech software to prepare the audio files. We adopted a female voice to ensure consistency with the avatars' appearance: Blessica for Tagalog (Filipino) and Mizuki for Japanese. Despite being different, the voices in the two languages were sufficiently similar to ensure that most of the participants did not notice either in pilot tests or in the main experiment. Instruction voices, such as announcements marking the beginning of word presentations, were also created using Ondoku.

\subsubsection{Virtual Environment}
The virtual environment was a room having a gray-stone floor, walls, and ceiling, consistent with a previous study~\cite{Mizuho-etal2024tvcg2}. We maintained the room as simple as possible to avoid unintentional confounding from other virtual objects and to naturally draw attention to the partner avatar.
Participants were positioned in the center of the room, with the partner avatar placed 1 m in front of them. This distance was determined through informal pilot tests and informed by previous research on proximity, according to which a distance of 0.7 to 2 m from a virtual human represents the boundary between comfortable and uncomfortable interactions~\cite{Bonsch-etal2018, Zibrek-etal2017, Zibrek-etal2018, Zibrek-McDonnell2019, Zibrek-etal2020}.

\subsubsection{Questionnaires}
We used three questionnaires: social presence survey (SPS), igroup presence questionnaire (IPQ), and simulator sickness questionnaire (SSQ). First, the SPS measured the subjective feeling of social presence of the virtual character and was adopted from previous studies by Bailenson et al.~\cite{Bailenson-etal2003}. It was also used by Zibrek et al. to examine virtual humans in IVEs~\cite{Zibrek-etal2019, Zibrek-McDonnell2019}. The SPS included five items rated on a scale from 1 (not at all) to 7 (extremely), with two of them being reversed. After inverting the ratings of the reversed items (e.g., 2 to 6), the average score of the five items was calculated as the representative score of perceived social presence. Participants answered the SPS at the end of each VR trial. We discuss the potential impact of social presence on memory performance.

Second, the IPQ measures the sense of presence~\cite{Schubert-etal2001}. It includes 14 questions, each answered on a seven-point Likert scale ranging from 1 to 7. After reversing the ratings of three inverted items, the average score of all items was used to represent perceived presence. Participants answered the IPQ once at the end of Day 1. This measure was used to assess the immersiveness of the VR experiences.

Finally, the SSQ measures the degree of cybersickness~\cite{Kennedy-etal1993}. It consists of 16 items, with each question answered on a four-point Likert scale ranging from 0 to 3. Three subscales (nausea, oculomotor, and disorientation) were calculated from the corresponding seven items, and then the total severity score was calculated based on these subscales. The maximum total severity score is 235.62. Participants answered the SSQ at the beginning and end of Day 1. We assessed whether prolonged VR exposure could induce severe cybersickness, which could confound differences in memory performance between conditions~\cite{Smith2019}.

\subsection{Measurements and Hypotheses}
The primary measurements were (1) acquisition, the proportion of correct responses on the five RP trials; (2) retention, the proportion of correct responses on the FT trial; and (3) forgetting, the proportion of forgetting from RP5 to FT. Forgetting was calculated using the following formula adopted from previous research~\cite{Mizuho-etal2024tvcg2}, where RP5 and FT are the scores on the corresponding tests:
\[Forgetting = \frac{FT - RP5}{RP5}\]

Retention and forgetting may appear identical. However, forgetting accounts for individual differences in the amount remembered at the end of Day 1 by incorporating the score from RP5 into the calculation. Previous studies suggest that the multiple-context effect is more evident in forgetting than in retention~\cite{Mizuho-etal2024tvcg1, Smith-Handy2016}. Therefore, examining both retention and forgetting requires investigation. Scoring was performed by reviewing the video recordings in the VR and real environments. Notably, as one participant reported that he originally knew two words, his proportion of correct responses was calculated by excluding words he knew from the total set of words, as in previous studies~\cite{Mizuho-etal2024tvcg2}. 

Regarding acquisition, in the constant avatar condition, the RP trials were conducted in the same social context as the IS, which enhanced recall due to the reinstatement effect. In contrast, in the varied avatar condition, the RP trials occurred in a different social context from the IS, which would reduce memory performance due to the reinstatement effect.
For retention, the varied avatar condition, where participants repeatedly learned in diverse social contexts, exhibited better memory retention than the constant avatar condition based on the multiple-context effect. Similarly, the varied avatar condition would show less forgetting than the constant avatar condition, again due to the multiple-context effect. Therefore, the following hypotheses were considered:
\begin{description}
    \item[H1] Acquisition under the constant avatar condition is better than that under the varied avatar condition.
    \item[H2] Retention under the varied avatar condition is higher than that under the constant avatar condition.
    \item[H3] Forgetting under the varied avatar condition is less than that under the constant avatar condition.
\end{description}

We also measured the social presence of partner avatars using SPS, presence in the IVE using IPQ, and cybersickness using SSQ.
It should be noted that these metrics were not intended to detect differences between avatar conditions, but rather to provide an overall understanding of the VR experience.
Additionally, we measured proximity during the avatar familiarization phase (\Cref{sec:procedure}) by recording the closest horizontal distance between the avatar (virtual character's position) and the participant (HMD position) in 30 seconds. Proximity served as a behavioral indicator of how participants perceived the partner avatars.
We also calculated correlations between social presence and memory performance in an exploratory manner to examine the mechanisms underlying the results.

\subsection{Procedures}
\label{sec:procedure}
\subsubsection{Day 1}
\Cref{fig:design} illustrates an overview of the experiment.
Participants were provided an overview of the experiment and signed a consent form. Subsequently, they received a detailed explanation of the experimental tasks. At this stage, they were notified of Day 1 procedures alone and received no details about Day 2, except for the time and location. This approach ensured that the final test on Day 2 was a surprise, similar to previous studies~\cite{Mizuho-etal2024tvcg1, Smith-Handy2016}, and helped minimize rehearsals during the one-week interval. Before undergoing the VR experience, participants answered the SSQ to assess their physical condition.

The participants performed six trials: IS and RP1--5. In the IS trial, they studied 20 Tagalog--Japanese pairs in the following steps: After wearing the HMD and headphones, the participants first went through a phase to familiarize themselves with the partner avatar. They were instructed to approach and observe the avatar for 30 seconds. This duration was consistent with the time used in previous studies using IVEs~\cite{Mizuho-etal2024tvcg1} or self-avatars~\cite{Mizuho-etal2024tvcg2} to enable participants' acclimatization to the stimuli. The participants could move around during this phase, and their positions were recorded to assess proximity. After resetting the avatar and participant positions to their initial values, they underwent a study phase. The partner avatar spoke Tagalog and Japanese in that order for 5 seconds; this was followed by a repetition of the pair within the next 5 seconds. This process was repeated for all 20 pairs in random order. Following the completion of the word presentation, the participants' HMD and headphones were removed, and they responded to the SPS.

Subsequently, the participants underwent the RP1 trial. First, they wore the HMD and headphones and observed the partner avatar for 30 seconds, as in the IS trial. Next, a Tagalog word was spoken, and they were asked to answer the corresponding Japanese translation orally within 5 seconds. Following this response time, they received feedback: The Tagalog word and its correct Japanese translation were spoken in sequence for an additional 5 seconds. This procedure was repeated for all 20 pairs in random order. Finally, after removing the HMD, participants answered the SPS.

The same RP trial was repeated four more times (RP2--5). Each trial was performed without any explicit breaks as in previous studies~\cite{Mizuho-etal2024tvcg1, Mizuho-etal2024tvcg2, Smith-Handy2014}.
After the RP5 trial, participants completed SSQ, IPQ, and a verbal interview. They were asked about their impressions of the realism of the avatar's appearance, behavior, and voice. The procedures for Day 1 lasted approximately 60 minutes.

\subsubsection{Day 2}
A week after Day 1, participants returned to a different room for the FT trial. The room was changed to avoid confounding arising from room matching~\cite{Smith1984}, which was consistent with previous studies~\cite{Mizuho-etal2024tvcg2}. The FT was performed in the physical world rather than the IVE, as in a previous study~\cite{Mizuho-etal2024tvcg2}. As mentioned earlier, the FT was an unexpected test for the participants. Before the test, they rated their confidence in answering the test on a scale from 0 to 100. They sat in front of a monitor and were instructed to focus on it during the test. They listened to a Tagalog word presented by a speaker and had 5 seconds to answer the corresponding Japanese translation. Once the response time elapsed, the next Tagalog word was presented. The FT did not include a feedback period. In this manner, 20 pairs were tested in random order. Notably, no avatar images were presented during the FT, but the voice uttering the words was identical to that of Day 1. 
After the test, participants were interviewed about their impressions regarding the experiment and then dismissed. Day 2 procedures lasted approximately 15 minutes.

\section{Results} 
One participant in the constant avatar condition scored 0 for RP1 and RP2, which were more than 2 \textit{SD} below the mean. Additionally, one participant in the varied avatar condition scored 65\% in RP1 and 90\% in RP2, which were more than 2 \textit{SD} above the mean. Hence, the learning curves of these participants largely deviated from those of the other participants.
Therefore, we excluded their data from the memory performance analyses as outliers.
However, as no technical issues occurred during the experiment, we decided to include them in the analyses of questionnaires and proximity.

\subsection{Acquisition}
\begin{figure*}[tb]
    \centering
    \subfloat[Recall]{\includegraphics[clip, width=\columnwidth]{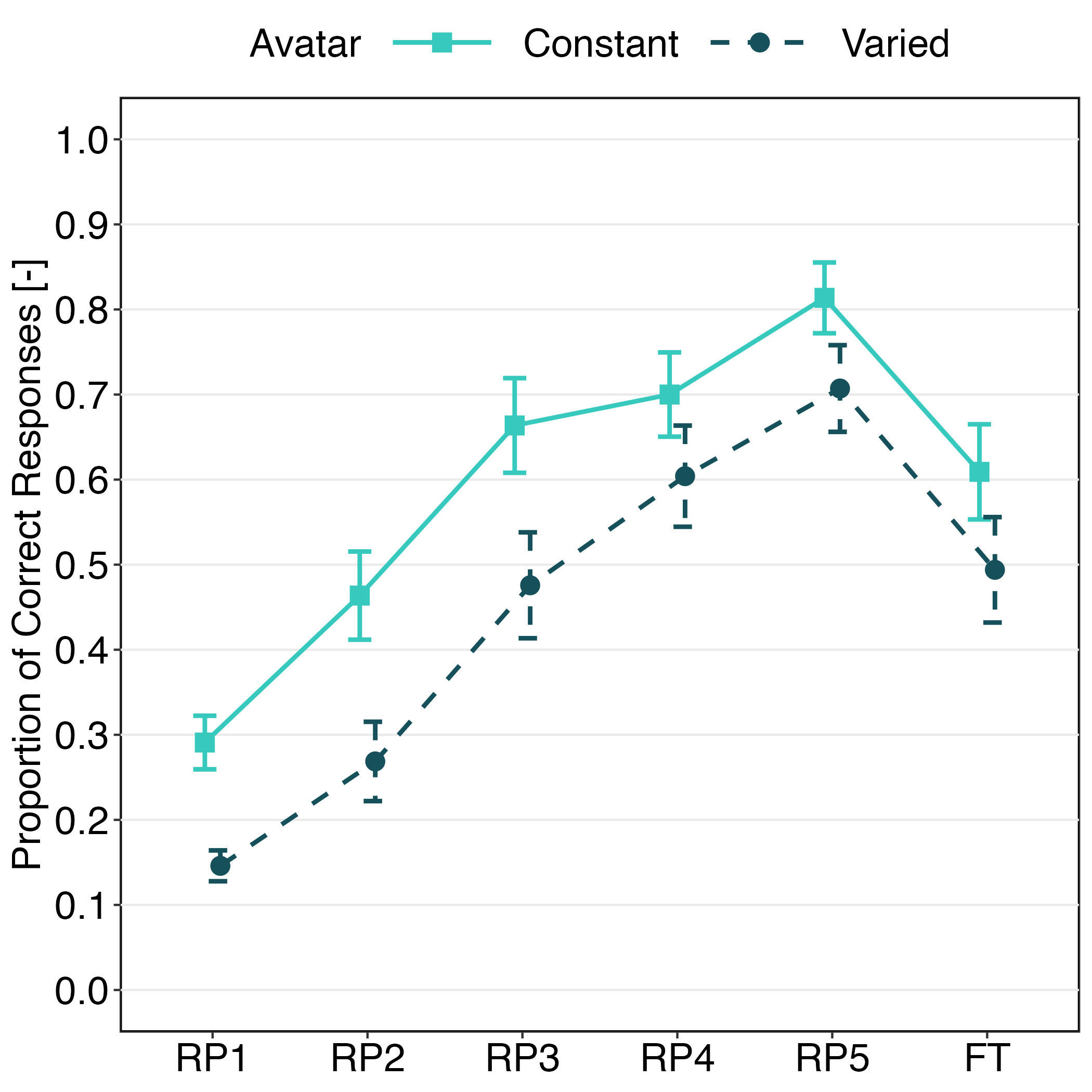}%
        \label{fig:result:recall}}
    \hfill%
    \subfloat[Forgetting]{\includegraphics[clip, width=\columnwidth]{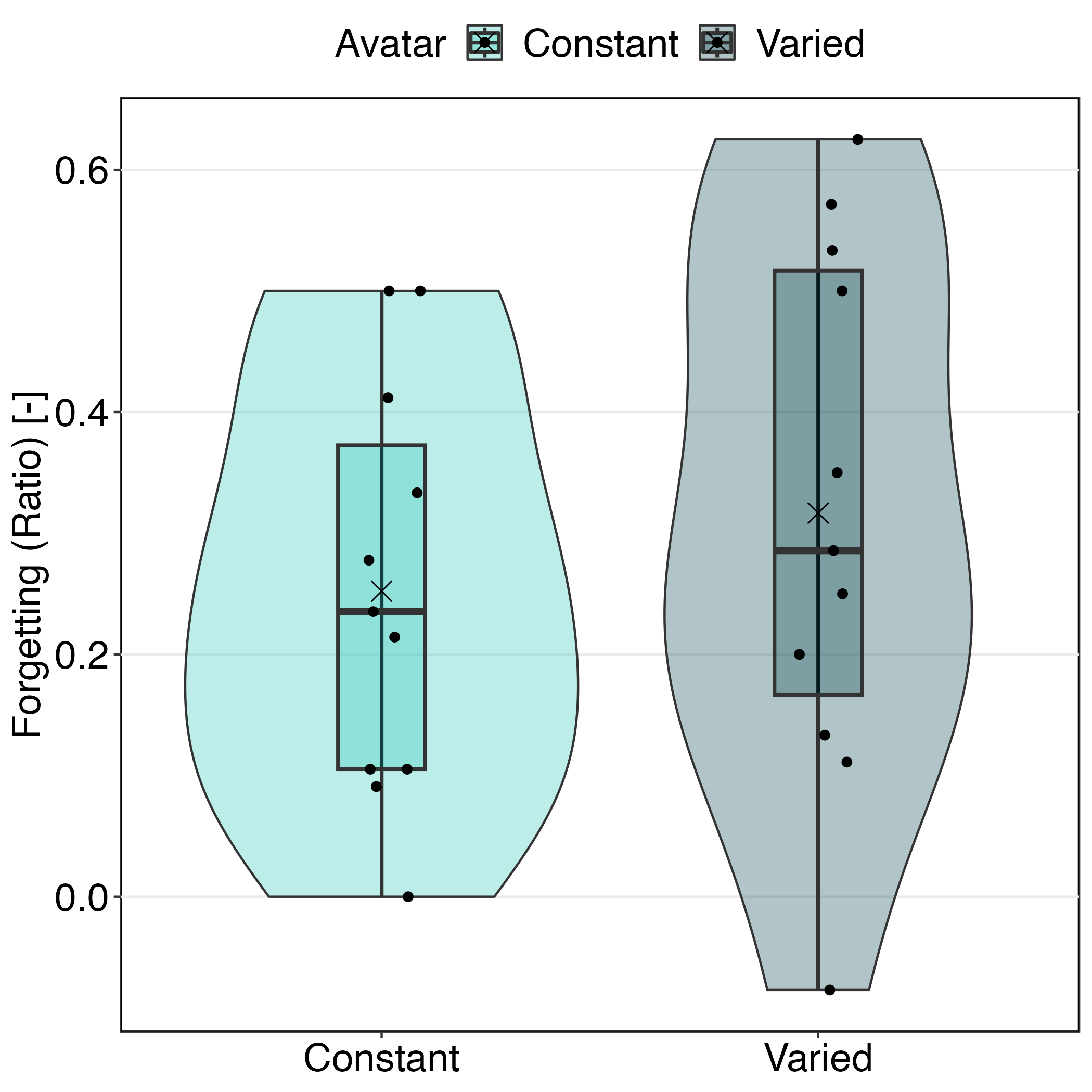}%
        \label{fig:result:forgetting}}
    \caption{(a) Proportion of correct responses for RP1--5 and FT. The constant avatar condition showed significantly better recall during RP1--5 than the varied avatar condition. Error bars represent standard errors. (b) Forgetting ratio from RP5 to FT. We found no significant differences between the conditions.}
    \label{fig:result:memory}
\end{figure*}

\Cref{fig:result:recall} shows the proportion of correct responses in RP1--5 and the FT. 
The normality assumption was not violated (Shapiro-Wilk test, \textit{p} $>$ .05); however, the sphericity assumption was violated (Mendoza's multi-sample sphericity test, \textit{p} = .014, Greenhouse-Geisser $\epsilon$ = 0.69). Therefore, we conducted an avatar (constant, varied) $\times$ trial (RP1--5) two-way analysis of variance (ANOVA) with Greenhouse-Geisser correction. The results showed a significant main effect of the avatar factor (\textit{F} (1, 20) = 5.95, \textit{p} = .024, ${\eta_p}^2$ = 0.23, $1 - \beta$ = 0.88) and the trial factor (\textit{F} (2.77, 55.4) = 138.8, \textit{p} $<$ .001, ${\eta_p}^2$ = 0.87, $1 - \beta$ = 1.00). However, we found no significant interaction effect of the two factors (\textit{F} (2.77, 55.4) = 1.483, \textit{p} = .231, ${\eta_p}^2$ = 0.07, $1 - \beta$ = 0.90). Therefore, the constant avatar condition exhibited better recall than the varied avatar condition in RP1-5 trials.
Regarding the trial's main effect, post-hoc tests using Shaffer's modified sequentially rejective Bonferroni method showed significant differences between any two levels ($p_{adj} < .01$).

\subsection{Retention}
As the normality (Shapiro-Wilk test, \textit{p} $>$ .05) and homogeneity of variance assumptions (F test, \textit{p} $>$ .05) were not violated, we performed a Student's \textit{t}-test. The results showed no significant difference between the constant and varied avatar conditions (\textit{t} (20) = 1.38, \textit{p} = 0.183, \textit{d} = 0.59, 1 - $\beta$ = 0.26).

\subsection{Forgetting}
\Cref{fig:result:forgetting} shows the forgetting ratios from RP5 to FT. As the normality (Shapiro-Wilk test, \textit{p} $>$ .05) and homogeneity of variance assumptions (F test, \textit{p} $>$ .05) were not violated, we performed a Student's \textit{t}-test. The results showed no significant difference between the constant and varied avatar conditions (\textit{t} (20) = 0.76, \textit{p} = 0.453, \textit{d} = 0.33, 1 - $\beta$ = 0.11).

\subsection{Questionnaires}
\subsubsection{Social Presence Survey (SPS)}
\begin{figure*}[tb]
    \centering
    \begin{tabular}{c}
		\begin{minipage}[c]{0.3\hsize}
	    \centering
        \subfloat[Social Presence]{\includegraphics[clip, width=\columnwidth]{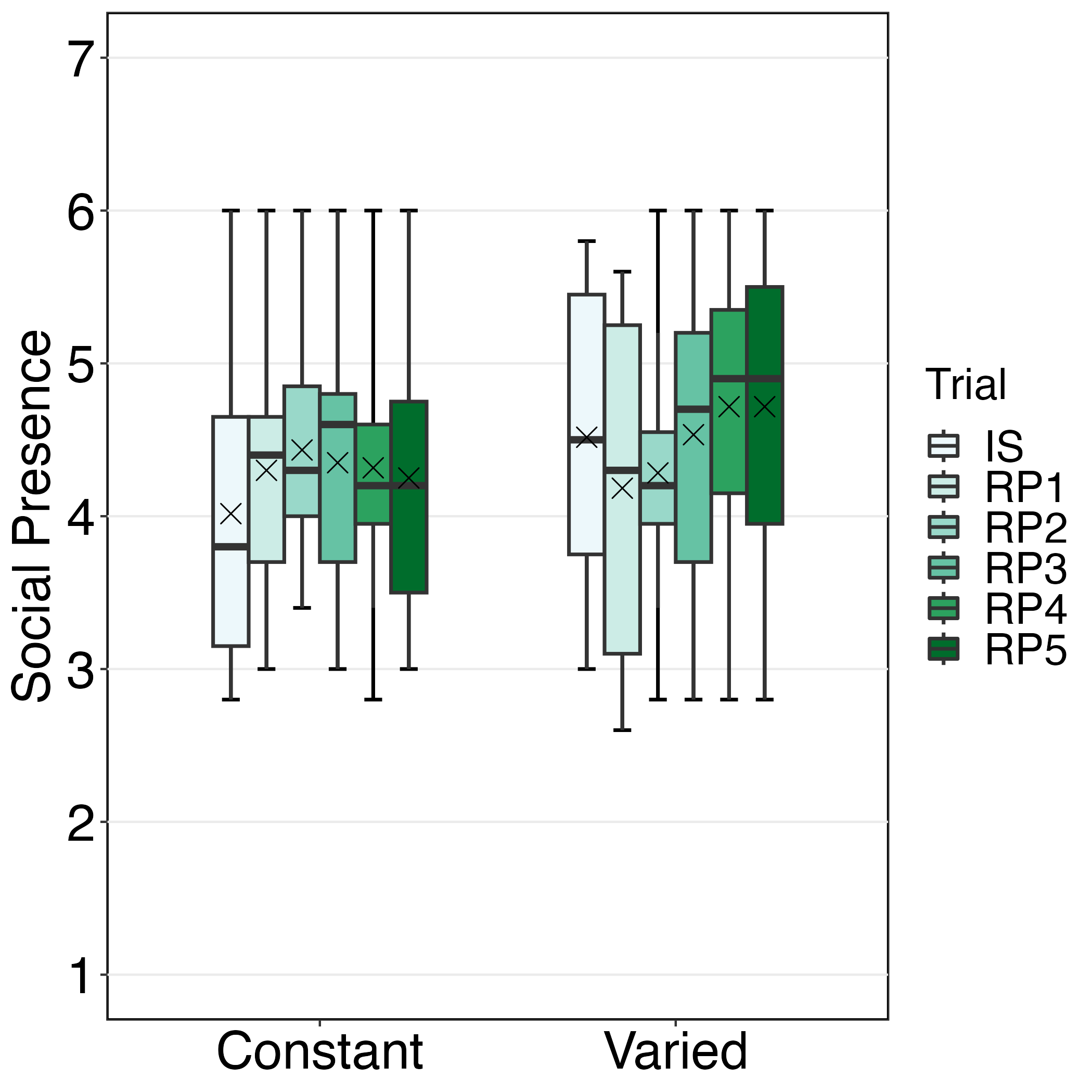}%
            \label{fig:result:social}}
        \end{minipage}
        \hspace{5pt}
        \begin{minipage}[c]{0.3\hsize}
	    \centering
        \subfloat[Presence]{\includegraphics[clip, width=\columnwidth]{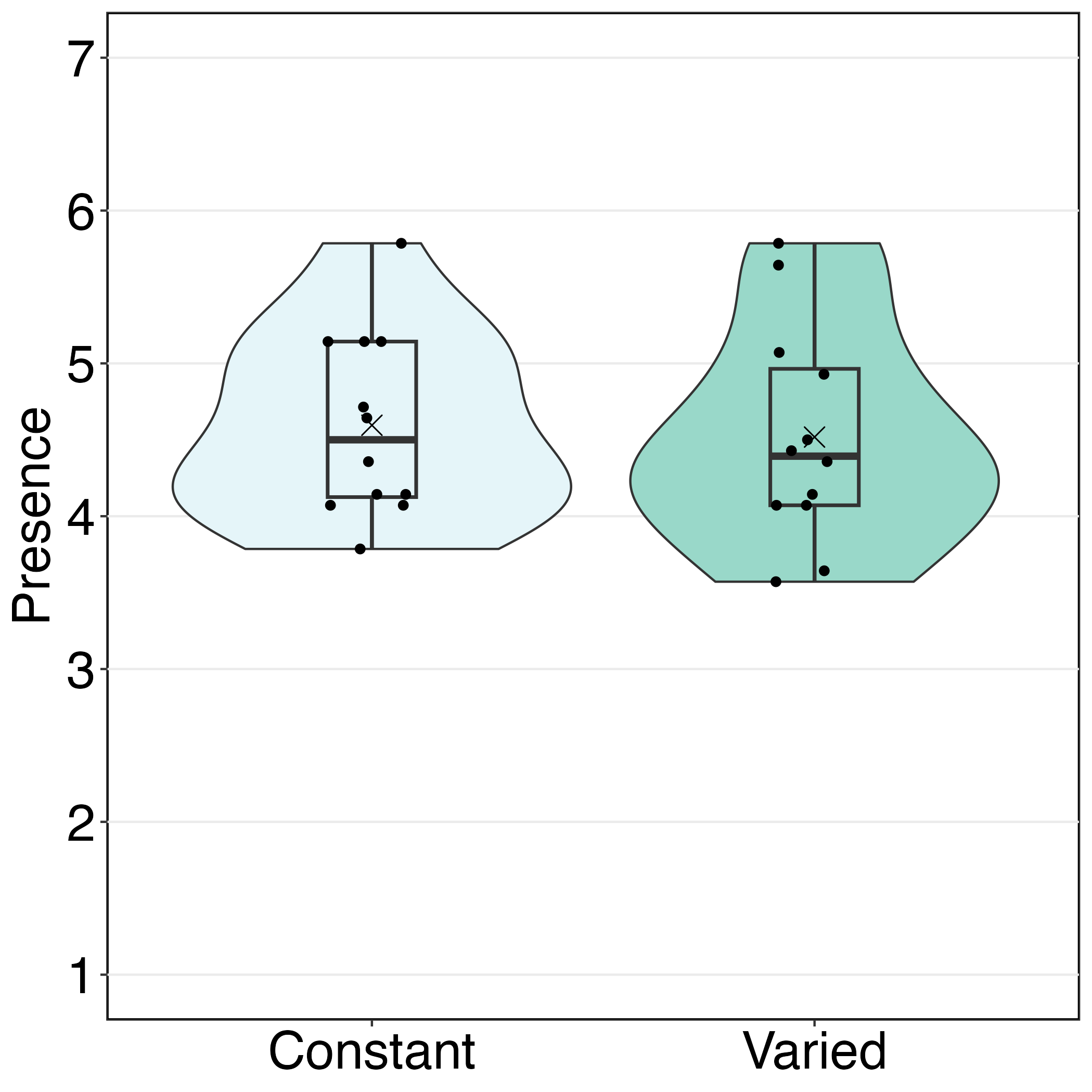}%
            \label{fig:result:ipq}}
        \end{minipage}
        \hspace{5pt}
        \begin{minipage}[c]{0.3\hsize}
	    \centering
        \subfloat[Cybersickness]{\includegraphics[clip, width=\columnwidth]{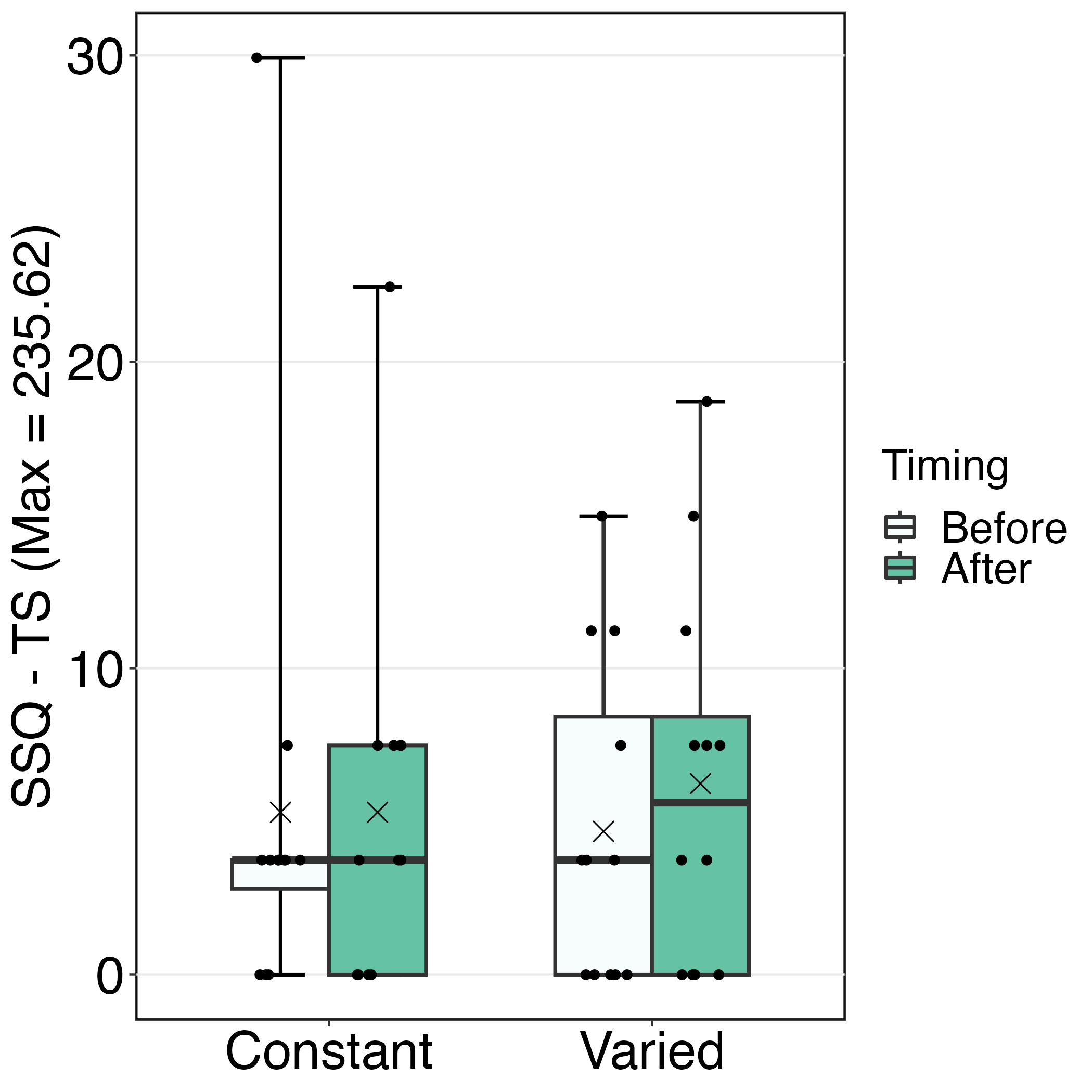}%
            \label{fig:result:ssq}}
        \end{minipage}
    \end{tabular}
    \caption{Questionnaire results. No significant differences were found between the conditions.}
    \label{fig:result:questionnaire}
\end{figure*}

\Cref{fig:result:social} shows the degree of social presence measured using SPS. We first performed an aligned rank transform (ART)~\cite{Wobbrock-etal2011} and conducted an avatar (constant, varied) $\times$ trial (IS, RP1--5) two-way ANOVA. ART enabled us to conduct an ANOVA for non-parametric data. The results showed no significant main effect of the avatar (\textit{F} (1, 22) = 0.40, \textit{p} = .536, ${\eta_p}^2$ = 0.02), main effect of the trial (\textit{F} (5, 110) = 0.99, \textit{p} = .430, ${\eta_p}^2$ = 0.04), or interaction effect of the two (\textit{F} (5, 110) = 1.72, \textit{p} = .136, ${\eta_p}^2$ = 0.07).

\subsubsection{Igroup Presence Questionnaire (IPQ)}
\Cref{fig:result:ipq} shows the degree of presence. We performed a Wilcoxon rank sum test, which showed no significant differences between the constant and varied avatar conditions (\textit{p} = .659, \textit{r} = 0.09).

\subsubsection{Simulator Sickness Questionnaire (SSQ)}
\Cref{fig:result:ssq} shows the degree of cybersickness. We first performed ART and then an avatar $\times$ timing (before or after the experiment) two-way ANOVA. The results showed no significant main effect of the avatar (\textit{F} (1, 22) = 0.09, \textit{p} = .770, ${\eta_p}^2$ = 0.00), main effect of the timing (\textit{F} (1, 22) = 0.86, \textit{p} = .364, ${\eta_p}^2$ = 0.04), or interaction effect of the two (\textit{F} (1, 22) = 0.16, \textit{p} = .695, ${\eta_p}^2$ = 0.01).

\subsubsection{Confidence}
Regarding their confidence in the FT, participants in the constant condition reported an average of 31.36 (\textit{SD} = 20.26) out of 100. In contrast, those in the varied condition reported an average of 29.27 (\textit{SD} = 21.74).
As the normality assumption was violated (Shapiro-Wilk test, \textit{p} $<$ .05), we conducted a Wilcoxon rank sum test, which showed no significant difference between the conditions (\textit{p} = .883, \textit{r} = 0.03).

\subsection{Proximity}
\begin{figure}[tb]
 \centering
 \includegraphics[width=\columnwidth]{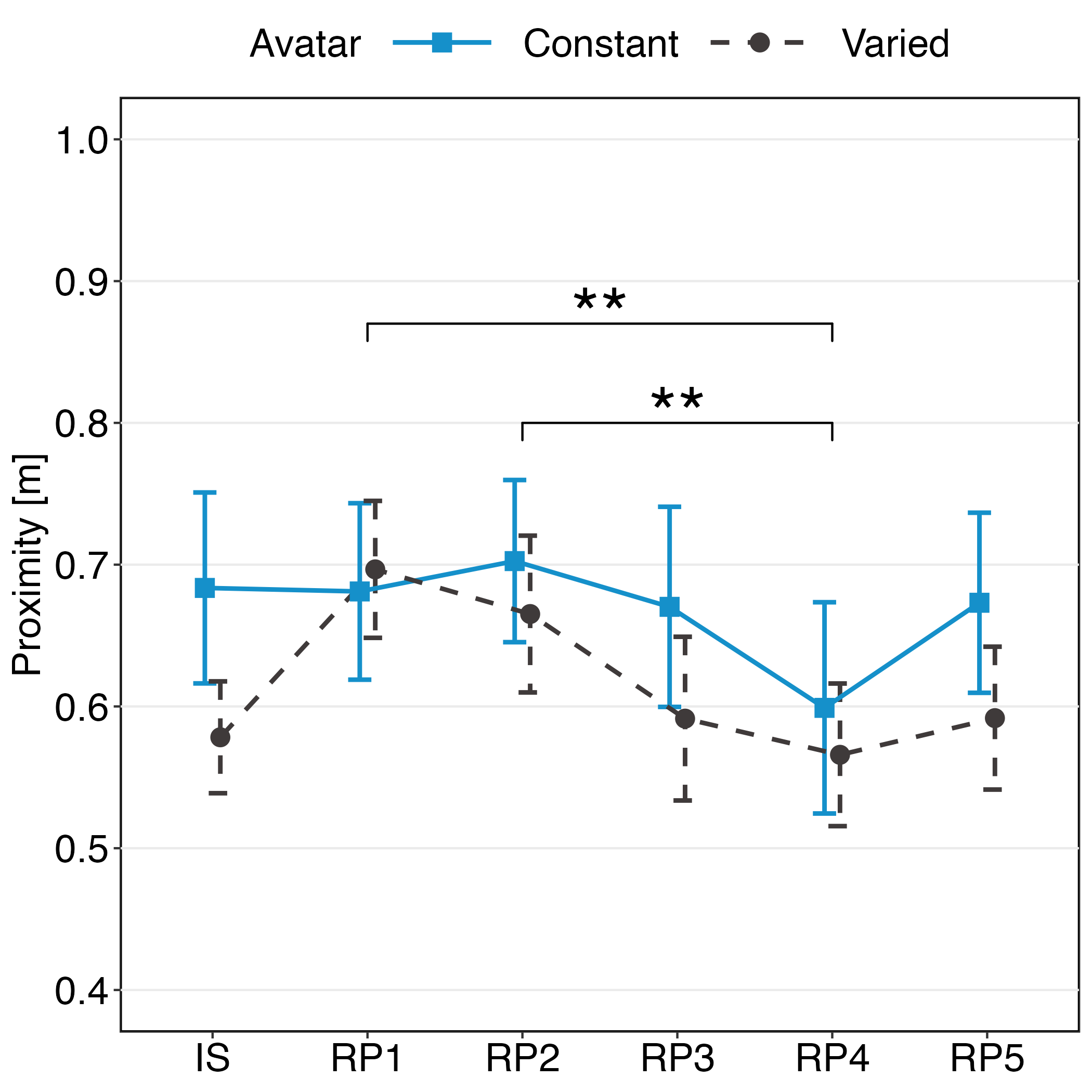}
 \caption{Proximity. Participants approached significantly closer to the partner avatar on RP4 than RP1 and RP2. Error bars represent the standard errors. $^{**}$: \textit{p} $<$ .01.}
 \label{fig:result:proximity}
\end{figure}

\Cref{fig:result:proximity} shows the minimum distance from the partner avatar during the avatar familiarization phase for each trial.
The normality assumption was not violated (Shapiro-Wilk test, \textit{p} $>$ .05). However, the sphericity assumption was violated (Mendoza's multi-sample sphericity test, \textit{p} = .004, Greenhouse-Geisser $\epsilon$ = 0.65). Therefore, we conducted an avatar (constant, varied) $\times$ trial (IS, RP1-5) two-way ANOVA with the Greenhouse-Geisser correction. The results showed a significant main effect of the trial factor (\textit{F} (3.24, 71.4) = 3.98, \textit{p} = .009, ${\eta_p}^2$ = 0.15). Post hoc tests under Shaffer's modified sequentially rejective Bonferroni method showed the distance in RP4 to be significantly shorter than that in RP1 (\textit{t} (22) = 3.13, $p_{adj}$ = .048) and RP2 (\textit{t} (22) = 3.82, $p_{adj}$ = .014).
No other pairs showed any significant difference ($p_{adj} > .05$). Additionally, the main effect of the avatar factor (\textit{F} (1, 22) = 0.51, \textit{p} = .484, ${\eta_p}^2$ = 0.02) and the interaction effect between the avatar and trial (\textit{F} (3.24, 71.4) = 1.21, \textit{p} = .312, ${\eta_p}^2$ = 0.05) were not significant.

\subsection{Correlation between Social Presence and Recall}
\begin{figure*}[tb]
    \centering
    \begin{tabular}{c}
		\begin{minipage}[c]{0.3\hsize}
	    \centering
        \subfloat[RP1]{\includegraphics[clip, width=\columnwidth]{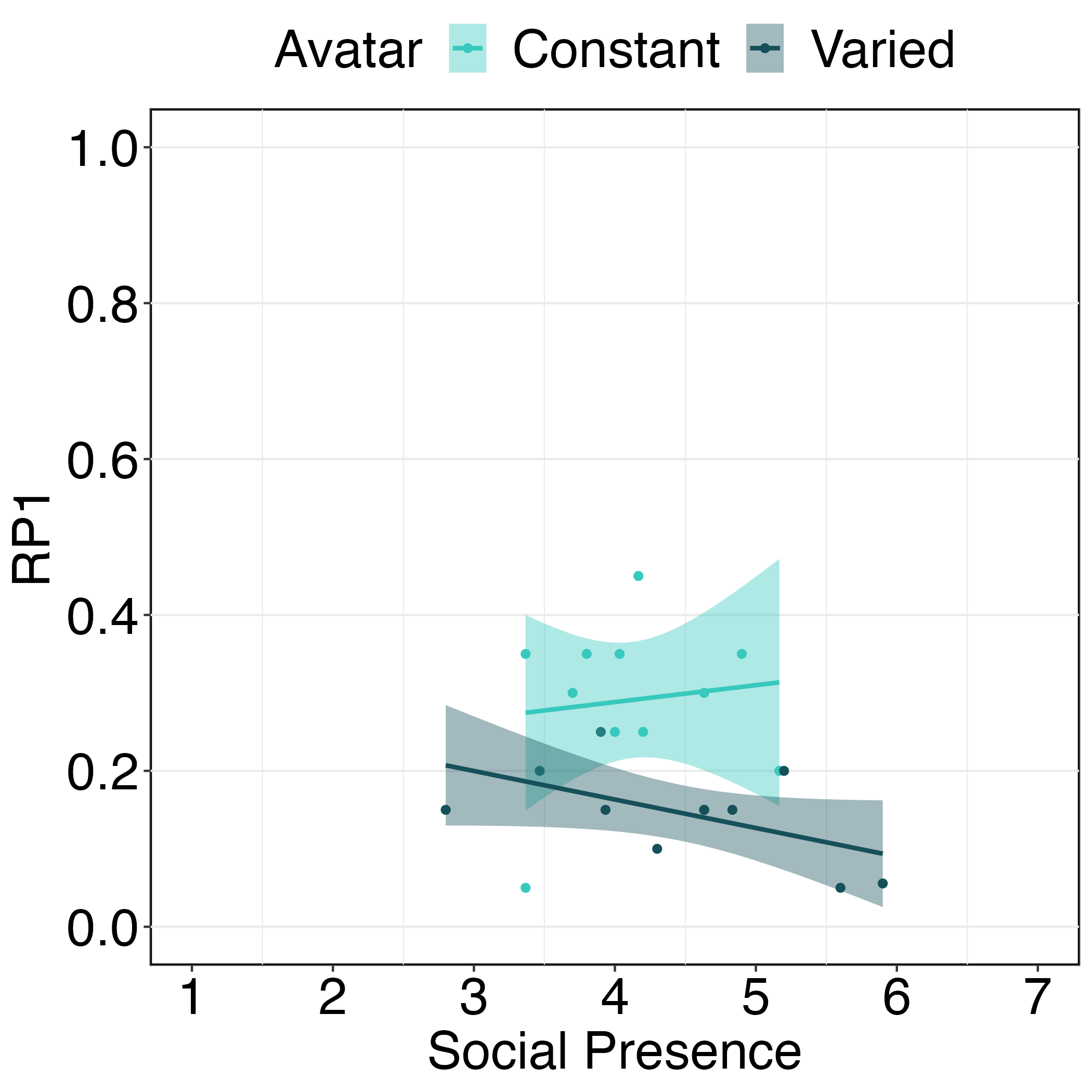}%
            \label{fig:result:rp1-social}}
        \end{minipage}    
        \hspace{5pt}
        \begin{minipage}[c]{0.3\hsize}
	    \centering
        \subfloat[FT]{\includegraphics[clip, width=\columnwidth]{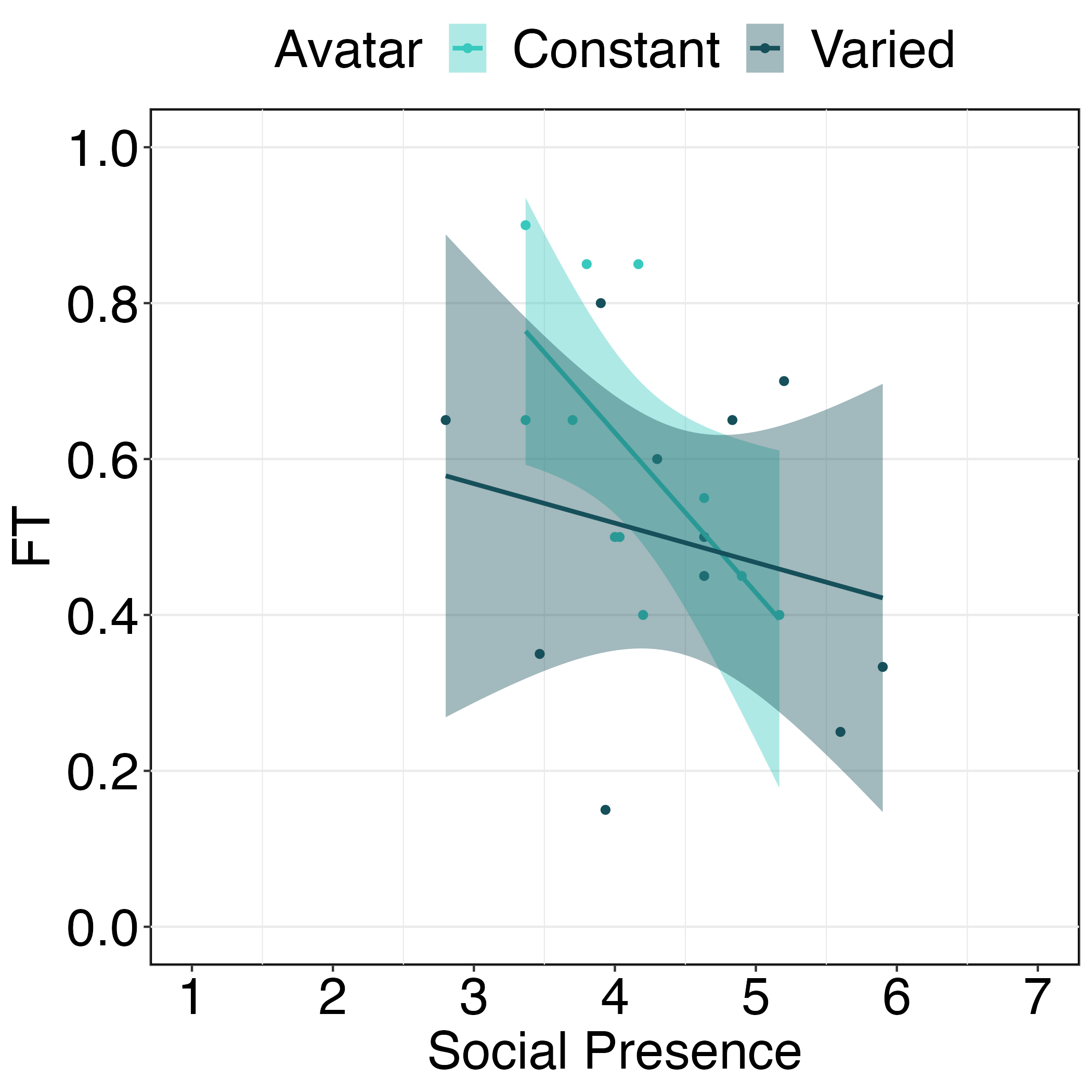}%
            \label{fig:result:ft-social}}
        \end{minipage}
        \hspace{5pt}
        \begin{minipage}[c]{0.3\hsize}
	    \centering
        \subfloat[Forgetting]{\includegraphics[clip, width=\columnwidth]{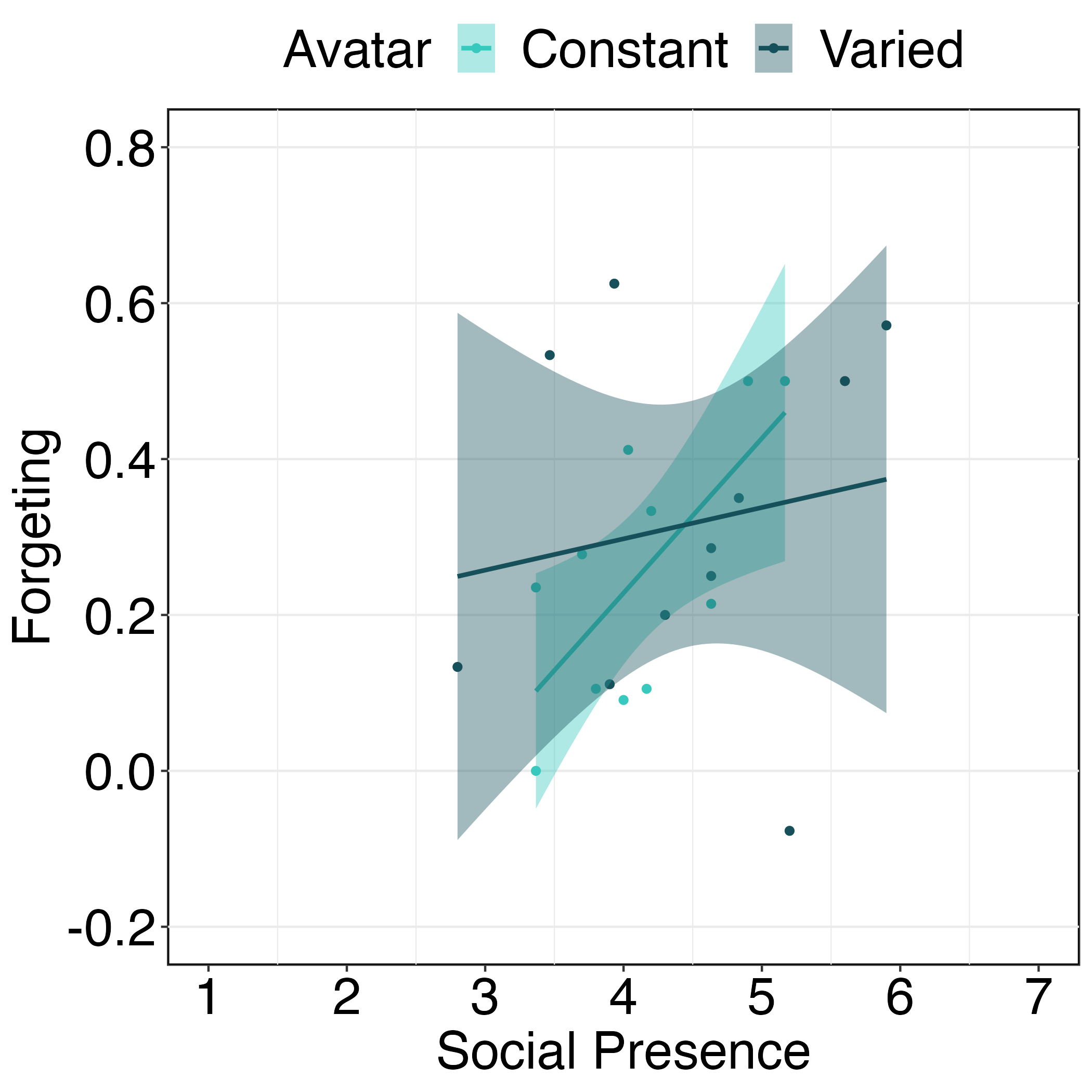}%
            \label{fig:result:forgetting-social}}
        \end{minipage}
    \end{tabular}
    \caption{Correlation between social presence and recall performance. Shaded regions indicate 95\% confidence intervals.}
    \label{fig:result:recall-social}
\end{figure*}

We calculated the correlation between social presence and recall performance.
As there was no difference in social presence between trials, we used the average of the six trials as the representative value. \Cref{fig:result:recall-social} shows the regression lines between social presence and RP1, FT, and Forgetting. Pearson's product-moment correlation analyses (Holm corrected) showed no significant correlation between RP1 and social presence in either the constant ($p_{adj}$ = .723, \textit{r} = 0.12) or varied condition ($p_{adj}$ = .147, \textit{r} = -0.56). Additionally, we found a marginally significant negative correlation between FT and social presence under the constant condition ($p_{adj}$ = .063, \textit{r} = -0.65) but not under the varied condition ($p_{adj}$ = .503, \textit{r} = -0.22). Finally, a significant positive correlation was found between Forgetting and social presence in the constant condition ($p_{adj}$ = .042, \textit{r} = 0.68) but not in the varied condition ($p_{adj}$ = .623, \textit{r} = 0.16).

\section{Discussion}

\subsection{Constant Avatar Led to Better Memory Acquisition}
As shown in \cref{fig:result:recall}, recall performance was significantly better in the constant avatar condition than in the varied avatar condition across RP1--5 trials. The effect size was large (${\eta_p}^2$ = 0.23), and the statistical power was appropriate (power = 0.88). This result supported hypothesis H1. In the constant condition, RP trials were conducted in the same social context as in learning. Hence, recall performance was high due to the match of environmental context. Conversely, in the varied condition, RP trials were conducted in a different social context each time, which diminished the availability of contextual cues and resulted in lower recall performance.

This study is the first to demonstrate the reinstatement effect of partner avatars in an IVE. Although Takenaka et al.~\cite{Takenaka-etal2024} were similarly motivated, they manipulated the scenery and partner avatars simultaneously, instead of separating these factors. In contrast, this study focused solely on manipulating the partner avatar's appearance.
The influence of social context on memory has not been thoroughly investigated, largely due to the difficulty of controlling the appearance and behavior of others in real-world environments.
From this perspective, this study provides new evidence suggesting that VR holds significant potential for exploring the role of social context in episodic memory.
While there is growing interest in research on pedagogical agents in VR, this study is innovative in its focus on the combination of instructor avatars across the learning and testing phases.

The reinstatement effect observed in this study had a large effect size. To facilitate the comparison between the current and previous studies, we specifically examined RP1 results. The recall scores in RP1 were \textit{M} = 0.29 (\textit{SD} = 0.10) in the constant condition and \textit{M} = 0.15 (\textit{SD} = 0.06) in the varied condition, yielding an effect size of \textit{d} = 1.70. This is much larger than the reinstatement effect reported by previous studies. For instance, a meta-analysis~\cite{Smith-Vela2001} calculated the effect size of the reinstatement effect as \textit{d} = 0.28. Additionally, Shin et al.~\cite{Shin-etal2021} used three-dimensional computer graphics-based IVEs to obtain an effect size of \textit{d} = 0.60, whereas Mizuho et al.~\cite{Mizuho-etal2024tvcg2} used self-avatars to obtain \textit{d} = 0.92. One reason for the large effect size revealed in this study might be the close involvement of the partner avatar in the memory task. The partner avatar acted as the instructor, and the Tagalog and Japanese words were presented audibly through the avatar's mouth. If the words had been presented visually, as in Mizuho et al.'s experiment~\cite{Mizuho-etal2024tvcg1}, the avatar might have merely served as an observer, and the effect might have been weaker. Moreover, Mizuho et al.~\cite{Mizuho-etal2024tvcg2}, who used a self-avatar, attributed the large reinstatement effect obtained in their study to the application of a sign-language learning task, which involved physical movements and naturally drew attention to the self-avatar. Hence, a comprehensive investigation is required to determine whether the findings can be generalized to other memory materials.

Alternatively, partner avatars may exert a more substantial influence on the environmental context dependency of memory than other factors. For example, Smith and Vela~\cite{Smith-Vela2001} suggest that the reinstatement effect is more pronounced when both the place and experimenter are simultaneously manipulated than when the place alone is manipulated. Additionally, Isarida and Isarida~\cite{Isarida-Isarida2010} coined the term ``complex-place context'' to describe the simultaneous manipulation of place and social context (experimenters or the number of participants taking the experiment together). They suggest that this approach yields a more robust reinstatement effect than the case in which the place alone is manipulated. However, these studies do not specifically examine the effects of manipulating the social context alone. Hence, it is unclear whether the robust reinstatement effect resulted from the complex manipulation of both place and social context or the manipulation of social context alone. This gap in understanding reflects the difficulty of investigating social context's effect in the physical world. Therefore, future VR research should provide the means to explore and clarify this possibility.

To examine potential improvements in the reinstatement effect, we performed a correlation analysis between social presence and RP1 (\Cref{fig:result:rp1-social}). We found no significant correlation; however, considering the magnitude of the correlation coefficient may be valuable for future research. The correlation suggested a weak positive value in the constant condition and a medium to large negative value in the varied condition. This result might indicate that high social presence could amplify the difference between constant and varied conditions, thereby enhancing the reinstatement effect. Although this study could not provide a definitive conclusion due to limitations in statistical power, future studies on the relationship between social presence and recall would provide important insights.

\subsection{Lack of Significant Differences in Memory Retention}
As shown in \cref{fig:result:recall}, memory retention during the FT did not significantly differ between avatar conditions. Similarly, the magnitude of forgetting, shown in \cref{fig:result:forgetting}, showed no significant difference between the conditions. Hence, the results did not support H2 and H3, indicating that the multiple-context effect was not observed for partner avatar use in the IVE.

Researchers have examined the VR-based multiple-context effect using IVEs and self-avatars. In Mizuho et al.'s study using IVEs, the effect size on forgetting was \textit{d} = 0.81~\cite{Mizuho-etal2024tvcg1}; further, in their study using self-avatars, it was \textit{d} = 1.49~\cite{Mizuho-etal2024tvcg2}. In contrast, the effect size in this study was \textit{d} = 0.33. Furthermore, the superiority of the conditions contrasted with the hypotheses. The results suggested that the effects on memory may differ between IVEs/self-avatars and partner avatars. However, as the sample size in this study was determined based on the expectation of a large effect size and the observed effect size was small, the statistical power was insufficient (1 - $\beta \approx$ 0.2). Therefore, further study is required to verify these findings.
The qualitative difference between partner avatars and IVEs/self-avatars in their impact on the multiple-context effect may be related to the findings of Isarida and Isarida~\cite{Isarida-Isarida2010}, who compared simple-place context manipulation (in which only the place context changed) and complex-place context manipulation (in which both the place and social contexts changed). Results showed that the varied context condition resulted in better recall performance than the constant condition in simple-place context manipulation. However, the opposite occurred in the complex-place context manipulation, with better recall occurring in the constant, rather than varied, condition. Moreover, the multiple-context effect that occurred in simple-place contexts disappeared when the interval between learning sessions was changed, whereas the effect in complex-place contexts remained stable. These findings suggest that memory processing differs between simple- and complex-place contexts. Isarida and Isarida attributed this qualitative difference to the fact that simple-place context manipulation changes part of the experience while complex-place context manipulation shifts the entire experience. However, they did not isolate the effect of social context separately, leaving the mechanism unclear. These aspects have not been clarified due to the high cost of preparing multiple environmental contexts to examine the multiple-context effect and the difficulty of manipulating social context. As VR can resolve both these issues, it will accelerate such investigations.

The relationship between the multiple-context effect and social presence revealed an interesting pattern (\Cref{fig:result:forgetting-social}). We found a significant positive correlation between forgetting and social presence under the constant avatar condition; however, this was not observed under the varied avatar condition.
This trend suggests that in the constant condition, the higher the perceived social presence, the stronger the influence of the avatar as a contextual cue and, thereby, the greater context-dependent forgetting from VR to real. This finding implies that social presence is critical in enhancing the multiple-context effect. Hence, one potential avenue for future research is to explore how increasing social presence will enhance the hypothesized multiple-context effect. For example, while the avatar unilaterally presented the information in this study, the use of more interactive partner avatars can heighten social presence~\cite{Oh-etal2018}.

\subsection{How Participants Perceived Their Partner Avatars and VR Experiences}
As shown in \cref{fig:result:questionnaire}, no significant differences were found between the conditions for social presence, presence, or cybersickness. The average scores for social presence and presence were above 4 on a 7-point scale, indicating the perception of a certain degree of presence. On the other hand, cybersickness scores did not show any significant change before and after the experiment. As the maximum score on the scale was 235.62, the scores recorded were generally low. Therefore, it was unlikely that cybersickness confounded differences in memory performance between conditions.

There was no difference in FT confidence between avatar conditions, and participants in both conditions estimated their performance to be approximately 30\% lower than their actual performance. This tendency was also observed in a previous study~\cite{Mizuho-etal2024tvcg2}. The participants' inability to accurately estimate their recall ability should be examined in educational settings.

Further, there was no significant difference in proximity between avatar conditions. However, participants approached the avatar more closely in RP4 than in RP1 or RP2. This result aligns with the participants' feedback, for instance, ``I (gradually) began to think that the avatar was not human.'' Additionally, many participants mentioned the eeriness associated with the repetitive actions and lines of the avatar. Thus, repeated exposure to avatars reduced perceived humanness.
Although no significant effect on memory was observed in this study, the decline in perceived humanness may influence users' attitudes toward listening to avatars, which could eventually impact their memory performance.

In the verbal interviews, participants also provided feedback on the avatar's voice. Overall, it was perceived as a machine-generated voice. However, many participants noted that it felt familiar, as it resembled voices commonly encountered in English listening materials and smartphone voice assistants, making it easy to listen to. While the voice was not perceived as human, it was not considered unnatural. As previous research has examined the effects of the match or mismatch between the realism of an avatar's appearance and voice~\cite{Higgins-etal2022}, the role of voice in influencing user perception and memory remains an interesting topic for future exploration.

\section{Limitations and Future Work}
The sample size was appropriate for detecting the reinstatement effect but insufficient for clarifying the multiple-context effect. This limitation arose from the unexpectedly small and contrary effect size. Additionally, although the gender balance among participants was good, the sample was relatively young. Although no specific restrictions were mentioned during recruitment, the requirement for participants to visit the university twice, with a one-week interval between visits, probably limited the sample. For example, conducting remote experiments for participants owing HMDs, similar to the approach used by Mizuho et al.~\cite{Mizuho-etal2024tvcg1}, may attract a more diverse participant pool. However, this method can introduce bias toward those familiar with VR. Future research should address sample bias carefully to improve the generalizability.

As we employed a between-participants design, it could be argued that the observed differences between conditions reflect pre-existing differences in memory ability between groups.
While we cannot completely rule out this possibility, we believe it does not significantly undermine the study's contribution, given that (1) the experimental design aligns with the typical paradigms used by Smith and Handy~\cite{Smith-Handy2014, Smith-Handy2016} and Mizuho et al.~\cite{Mizuho-etal2024tvcg1, Mizuho-etal2024tvcg2}, and (2) participants were randomly assigned to each condition, with consideration given to factors such as age and gender, in order to ensure as much homogeneity between groups as possible.
In future research, approaches such as measuring baselines or adopting within-participant designs could be considered. However, it is important to recognize that these methods may also introduce confounding factors, such as practice effects.

An alternative explanation for the observed differences between conditions is that the repeated switching of avatars in the varied avatar condition may have disrupted participants’ attention, thereby hindering learning, rather than reflecting context-dependent effects on episodic memory.
Indeed, previous studies have noted that pedagogical agents and avatars distract attention~\cite{Petersen-etal2021, Gamage-Ennis2018}. However, given that there was no difference in the recall performance in the FT, we believe the amount of information encoded did not differ between conditions. Instead, the results can be interpreted as the recall process being facilitated or hindered depending on whether the environmental context during RP trials was consistent or inconsistent with the IS. Notably, to minimize such confounding factors, we incorporated a familiarization phase before the memory task to allow participants to get accustomed to the partner avatar. In future studies, we can enhance our understanding of user attention by measuring task load and gaze, enabling a more detailed discussion of its impact on memory.

It is also possible that the avatars used in the constant avatar condition had varying effects on memory performance. For example, Beege et al.~\cite{Beege-etal2022} suggested that screen-based agents dressed in suits and speaking in a professional manner have a greater learning effect than agents dressed casually. Similarly, Makransky et al.~\cite{Makransky-etal2019JCAL} demonstrated that male students benefited more from a drone agent, while female students showed better learning outcomes with a female researcher agent, highlighting individual differences in effective agent design.
While differences in the avatars' clothing and age were necessary to vary the social context in this study, these characteristics may have introduced subtle confounds in memory performance. For instance, during our validation and informal pilot testing, some participants reported that the elderly avatar (Business\_Female\_02) was familiar and plausible as a lecturer, whereas the military avatar (Military\_Female\_02) was perceived as frightening, suggesting that the avatars elicited distinct emotional responses.
In this study, however, only two participants were assigned to each avatar in the constant avatar condition, making it difficult to determine the specific effect of each avatar.
Since the types and order of avatars used were counterbalanced between participants, we believe this limitation is unlikely to impact the overall research results significantly. Nevertheless, further investigation into the effects of individual avatars is necessary to gain deeper insights.

In this study, all the avatars were female in conformance with the trend observed in previous studies~\cite{Lin-etal2023, Patotskaya-etal2023, Zibrek-etal2018}, which typically used avatars of one gender. Ideally, avatars of both genders should be included to determine whether the avatar's gender influences memory. Although the results of this study are expected to be similar to the case with male avatars, further investigation is required.
An additional avenue for research is the effect of gender pairing between the participant and the partner avatar.
Additionally, avatars can vary widely beyond gender, including photorealistic, animal, or robot avatars~\cite{Mizuho-etal2023ahs, Mizuho-etal2024sap, Mizuho-etal2024frontiers}. Introducing such diverse avatars will highlight the differences between avatars and potentially lead to a stronger context-dependent effect. However, it should be noted that non-human avatars may exhibit lower social presence~\cite{Takenaka-etal2024} and result in less effective learning~\cite{Mizuho-etal2024sap} than human avatars.
Additionally, similarity in appearance between the self-avatar and the partner avatar has various effects on cognition~\cite{Latoschik-etal2017, Hu-etal2023}, and the combination of self and other avatars may alter the social context and thereby affect memory.
Besides appearance,  the effects of the avatars' movements, voice, and agency are also interesting.
Among these, the influence of agency (i.e., human-controlled vs. computer-controlled) is a particularly interesting topic, as existing findings on whether it contributes to the social impact of virtual entities are mixed and inconclusive (see the review by Oh et al.~\cite{Oh-etal2018} and Kyrlitsias and Michael-Grigoriou~\cite{Kyrlitsias-MichaelGrigoriou2022}). Perceived agency may alter the listener’s social attitude toward the entity and potentially affect memory in different ways.

It is also crucial to consider the generalizability of memory tasks. In this study, the partner avatar spoke Tagalog and Japanese words. We may obtain different outcomes if the avatar is not involved in the task, as seen in the study by Kruse et al.~\cite{Kruse-etal2023SUI}. The environmental context dependency of memory has primarily been studied in declarative (verbal) memory~\cite{Smith-Vela2001}. Therefore, while these findings likely apply to verbal memory, their effectiveness in motor skills learning remains uncertain. Additionally, various experimental designs should be considered to ensure generalizability, such as using different interval lengths between repeated learning, varying the time between the learning and final test (e.g., 5 minutes~\cite{Smith-Handy2014} or 1 month), conducting a final test in VR rather than in the real world, or comparing the effectiveness of partner avatars to those of the real human or sound-only conditions~\cite{Rzayev-etal2019, Petersen-etal2021}.

\section{Conclusion}
In this study, we examined the effects of partner avatars in an IVE on memory. Participants listened to 20 Tagalog--Japanese word pairs from a lecturer avatar and repeatedly studied them. Based on the environmental context dependency of episodic memory, we hypothesized that changing the partner avatar's appearance for each repetition (varied avatar condition) would lead to reduced recall performance during repetition compared to the use of a constant avatar (constant avatar condition). In contrast, we hypothesized that memory retention in the final test conducted in the real world one week later would be higher under the varied avatar condition. The results showed that the recall performance during repeated learning was significantly higher in the constant avatar condition than in the varied avatar condition, demonstrating for the first time the reinstatement effect of partner avatars in IVEs. Conversely, we found no difference between the conditions in memory retention in the final test. These findings suggest that memory processing with partner avatars may differ from that with IVEs or self-avatars. Additionally, we measured the social presence of partner avatars and explored its correlation with memory performance. The results suggested that higher social presence might enhance the multiple-context effect, suggesting the potential benefits of future research with increased social presence. To this day, the effects of instructor avatars in IVEs on memory remain underexplored. Therefore, we expect this study to serve as a foundation for further research into avatar designs that enhance the effectiveness of VR-based learning and training.

\section*{Acknowledgments}
This work was partially supported by
JST Moonshot Research \& Development (JPMJMS2013);
COI-NEXT (JPMJPF2201);
JSPS Grant-in-Aid for Challenging Research (Exploratory) (21K19784);
Grant-in-Aid for Research Activity Start-up (25K24404);
Grant-in-Aid for JSPS Fellows (23KJ0422);
and ``Multimodal XR-AI platform development for tele-habitation and reciprocal care coupling with health guidance'' project (JPNP21004) subsidized by NEDO.

The authors used OpenAI ChatGPT-3.5 for proofreading and language editing to improve the clarity and readability of the manuscript.

\bibliographystyle{IEEEtran}
\bibliography{reference}

@incollection{Bailenson-Blascovich2004,
    author = {Bailenson, Jeremy N. and Blascovich, Jim},
    title = {Avatars},
    booktitle = {Encyclopedia of human-computer interaction},
    publisher = {Berkshire Publishing Group},
    year = {2004}
}

@article{Kyrlitsias-MichaelGrigoriou2022,
  title = {Social {{Interaction With Agents}} and {{Avatars}} in {{Immersive Virtual Environments}}: {{A Survey}}},
  shorttitle = {Social {{Interaction With Agents}} and {{Avatars}} in {{Immersive Virtual Environments}}},
  author = {Kyrlitsias, Christos and {Michael-Grigoriou}, Despina},
  year = 2022,
  month = jan,
  journal = {Frontiers in Virtual Reality},
  volume = {2},
  publisher = {Frontiers},
  issn = {2673-4192},
  doi = {10.3389/frvir.2021.786665},
  urldate = {2026-02-05},
  langid = {american}
}

@inproceedings{Amemiya-etal2022,
  title = {Effect of~{{Face Appearance}} of~a~{{Teacher Avatar}} on~{{Active Participation During Online Live Class}}},
  booktitle = {Human {{Interface}} and the {{Management}} of {{Information}}: {{Applications}} in {{Complex Technological Environments}}},
  author = {Amemiya, Tomohiro and Aoyama, Kazuma and Ito, Kenichiro},
  editor = {Yamamoto, Sakae and Mori, Hirohiko},
  year = {2022},
  series = {Lecture {{Notes}} in {{Computer Science}}},
  pages = {99--110},
  publisher = {Springer International Publishing},
  address = {Cham},
  doi = {10.1007/978-3-031-06509-5_7},
  isbn = {978-3-031-06509-5},
  langid = {english}
}

@article{Aseeri-Interrante2021,
  title = {The {{Influence}} of {{Avatar Representation}} on {{Interpersonal Communication}} in {{Virtual Social Environments}}},
  author = {Aseeri, Sahar and Interrante, Victoria},
  year = {2021},
  month = may,
  journal = {IEEE Transactions on Visualization and Computer Graphics},
  volume = {27},
  number = {5},
  pages = {2608--2617},
  issn = {1941-0506},
  doi = {10.1109/TVCG.2021.3067783},
  urldate = {2023-11-30}
}

@article{Bailenson-etal2003,
  title = {Interpersonal {{Distance}} in {{Immersive Virtual Environments}}},
  author = {Bailenson, Jeremy N. and Blascovich, Jim and Beall, Andrew C. and Loomis, Jack M.},
  year = {2003},
  month = jul,
  journal = {Personality and Social Psychology Bulletin},
  volume = {29},
  number = {7},
  pages = {819--833},
  publisher = {SAGE Publications Inc},
  issn = {0146-1672},
  doi = {10.1177/0146167203029007002},
  urldate = {2024-04-06},
  langid = {english}
}

@inproceedings{Baylor-Kim2004,
  title = {Pedagogical {{Agent Design}}: {{The Impact}} of {{Agent Realism}}, {{Gender}}, {{Ethnicity}}, and {{Instructional Role}}},
  shorttitle = {Pedagogical {{Agent Design}}},
  booktitle = {Intelligent {{Tutoring Systems}}},
  author = {Baylor, Amy L. and Kim, Yanghee},
  editor = {Lester, James C. and Vicari, Rosa Maria and Paragua{\c c}u, F{\'a}bio},
  year = {2004},
  series = {Lecture {{Notes}} in {{Computer Science}}},
  pages = {592--603},
  publisher = {Springer},
  address = {Berlin, Heidelberg},
  doi = {10.1007/978-3-540-30139-4_56},
  isbn = {978-3-540-30139-4},
  langid = {english}
}

@article{Beege-etal2022,
  title = {How Instructors Influence Learning with Instructional Videos - {{The}} Importance of Professional Appearance and Communication},
  author = {Beege, Maik and Krieglstein, Felix and Arnold, Caroline},
  year = {2022},
  month = aug,
  journal = {Computers \& Education},
  volume = {185},
  pages = {104531},
  issn = {0360-1315},
  doi = {10.1016/j.compedu.2022.104531},
  urldate = {2022-08-19},
  langid = {english}
}

@incollection{Bjork-Bjork1992,
  title = {A New Theory of Disuse and an Old Theory of Stimulus Fluctuation},
  booktitle = {From Learning Processes to Cognitive Processes: {{Essays}} in Honor of {{William K}}. {{Estes}}},
  author = {Bjork, Robert A and Bjork, Elizabeth Ligon},
  editor = {Healy, A. and Kosslyn, S. and Shiffrin, R.},
  year = {1992},
  volume = {2},
  pages = {35--67},
  publisher = {Hillsdale, NJ: Erlbaum}
}

@inproceedings{Bonsch-etal2018,
  title = {Social {{VR}}: {{How Personal Space}} Is {{Affected}} by {{Virtual Agents}}' {{Emotions}}},
  shorttitle = {Social {{VR}}},
  booktitle = {2018 {{IEEE Conference}} on {{Virtual Reality}} and {{3D User Interfaces}} ({{VR}})},
  author = {B{\"o}nsch, Andrea and Radke, Sina and Overath, Heiko and Asch{\'e}, Laura M. and Wendt, Jonathan and Vierjahn, Tom and Habel, Ute and Kuhlen, Torsten W.},
  year = {2018},
  month = mar,
  pages = {199--206},
  doi = {10.1109/VR.2018.8446480},
  urldate = {2024-05-05}
}

@inproceedings{Bower1972,
  title = {Stimulus-Sampling Theory of Encoding Variability},
  booktitle = {Coding {{Processes}} in {{Human Memory}}},
  author = {Bower, Gordon H},
  editor = {Melton, Arthur W. and Martin, Edwin},
  year = {1972},
  pages = {85--123},
  publisher = {Winston},
  address = {Washington, D. C.}
}

@article{Chocholackova-etal2023,
  title = {Context-Dependent Memory Recall in {{HMD-based}} Immersive Virtual Environments},
  author = {Chochol{\'a}{\v c}kov{\'a}, M{\'a}ria and Ju{\v r}{\'i}k, Vojt{\v e}ch and Ru{\v z}i{\v c}kov{\'a}, Alexandra and Jurkovi{\v c}ov{\'a}, Lenka and Ugwitz, Pavel and Jel{\'i}nek, Martin},
  year = {2023},
  month = aug,
  journal = {PLOS ONE},
  volume = {18},
  number = {8},
  pages = {e0289079},
  publisher = {Public Library of Science},
  issn = {1932-6203},
  doi = {10.1371/journal.pone.0289079},
  urldate = {2023-08-29},
  langid = {english}
}

@inproceedings{DeBack-etal2018,
  title = {The {{Applicability}} and {{Benefits}} of {{Virtual Reality}} for the {{Cognitive Sciences}}: {{The Case}} of {{Context-Dependent Memory}}},
  shorttitle = {The {{Applicability}} and {{Benefits}} of {{Virtual Reality}} for the {{Cognitive Sciences}}},
  booktitle = {Proceedings of the 40th {{Annual Meeting}} of the {{Cognitive Science Society}}, {{CogSci}} 2018},
  author = {{de Back}, Tycho T. and Tinga, Angelica M. and {van Hoef}, Rens and Peters, Erwin M. and Louwerse, Max M.},
  year = {2018},
  month = jul,
  pages = {293--298},
  publisher = {The Cognitive Science Society},
  urldate = {2023-11-25},
  langid = {english}
}

@article{Faul-etal2007,
  title = {G*{{Power}} 3: {{A}} Flexible Statistical Power Analysis Program for the Social, Behavioral, and Biomedical Sciences},
  shorttitle = {G*{{Power}} 3},
  author = {Faul, Franz and Erdfelder, Edgar and Lang, Albert-Georg and Buchner, Axel},
  year = {2007},
  month = may,
  journal = {Behavior Research Methods},
  volume = {39},
  number = {2},
  pages = {175--191},
  issn = {1554-3528},
  doi = {10.3758/BF03193146},
  urldate = {2022-07-21},
  langid = {english}
}

@inproceedings{Gamage-Ennis2018,
  title = {Examining the Effects of a Virtual Character on Learning and Engagement in Serious Games},
  booktitle = {Proceedings of the 11th {{ACM SIGGRAPH Conference}} on {{Motion}}, {{Interaction}} and {{Games}}},
  author = {Gamage, Vihanga and Ennis, Cathy},
  year = {2018},
  month = nov,
  series = {{{MIG}} '18},
  pages = {1--9},
  publisher = {Association for Computing Machinery},
  address = {New York, NY, USA},
  doi = {10.1145/3274247.3274499},
  urldate = {2024-05-06},
  isbn = {978-1-4503-6015-9}
}

@article{Godden-Baddeley1975,
  title = {Context-Dependent Memory in Two Natural Environments: {{On}} Land and Underwater},
  author = {Godden, Duncan R and Baddeley, Alan D},
  year = {1975},
  journal = {British Journal of Psychology},
  volume = {66},
  number = {3},
  pages = {325--331},
  publisher = {Wiley Online Library},
  doi = {10.1111/j.2044-8295.1975.tb01468.x}
}

@inproceedings{GonzalezFranco-etal2020aivr,
  title = {{{MoveBox}}: {{Democratizing MoCap}} for the {{Microsoft Rocketbox Avatar Library}}},
  shorttitle = {{{MoveBox}}},
  booktitle = {2020 {{IEEE International Conference}} on {{Artificial Intelligence}} and {{Virtual Reality}} ({{AIVR}})},
  author = {{Gonzalez-Franco}, Mar and Egan, Zelia and Peachey, Matthew and Antley, Angus and Randhavane, Tanmay and Panda, Payod and Zhang, Yaying and Wang, Cheng Yao and Reilly, Derek F. and Peck, Tabitha C and Won, Andrea Stevenson and Steed, Anthony and Ofek, Eyal},
  year = {2020},
  month = feb,
  pages = {91--98},
  doi = {10.1109/AIVR50618.2020.00026},
  postfix = {aivr}
}

@article{GonzalezFranco-etal2020frontiers,
  title = {The {{Rocketbox Library}} and the {{Utility}} of {{Freely Available Rigged Avatars}}},
  author = {{Gonzalez-Franco}, Mar and Ofek, Eyal and Pan, Ye and Antley, Angus and Steed, Anthony and Spanlang, Bernhard and Maselli, Antonella and Banakou, Domna and Pelechano, Nuria and {Orts-Escolano}, Sergio and Orvalho, Veronica and Trutoiu, Laura and Wojcik, Markus and {Sanchez-Vives}, Maria V. and Bailenson, Jeremy and Slater, Mel and Lanier, Jaron},
  year = {2020},
  month = dec,
  journal = {Frontiers in Virtual Reality},
  volume = {1},
  issn = {2673-4192},
  doi = {10.3389/frvir.2020.561558},
  urldate = {2022-10-05},
  postfix = {frontiers}
}

@inproceedings{Guimaraes-etal2020,
  title = {The {{Impact}} of {{Virtual Reality}} in the {{Social Presence}} of a {{Virtual Agent}}},
  booktitle = {Proceedings of the 20th {{ACM International Conference}} on {{Intelligent Virtual Agents}}},
  author = {Guimar{\~a}es, Manuel and Prada, Rui and Santos, Pedro A. and Dias, Jo{\~a}o and Jhala, Arnav and Mascarenhas, Samuel},
  year = {2020},
  month = oct,
  series = {{{IVA}} '20},
  pages = {1--8},
  publisher = {Association for Computing Machinery},
  address = {New York, NY, USA},
  doi = {10.1145/3383652.3423879},
  urldate = {2024-09-13},
  isbn = {978-1-4503-7586-3}
}

@article{Higgins-etal2022,
  title = {Sympathy for the Digital: {{Influence}} of Synthetic Voice on Affinity, Social Presence and Empathy for Photorealistic Virtual Humans},
  shorttitle = {Sympathy for the Digital},
  author = {Higgins, Darragh and Zibrek, Katja and Cabral, Joao and Egan, Donal and McDonnell, Rachel},
  year = {2022},
  month = may,
  journal = {Computers \& Graphics},
  volume = {104},
  pages = {116--128},
  issn = {0097-8493},
  doi = {10.1016/j.cag.2022.03.009},
  urldate = {2023-11-25}
}

@article{Horovitz-Mayer2021,
  title = {Learning with Human and Virtual Instructors Who Display Happy or Bored Emotions in Video Lectures},
  author = {Horovitz, Tal and Mayer, Richard E.},
  year = {2021},
  month = jun,
  journal = {Computers in Human Behavior},
  volume = {119},
  pages = {106724},
  issn = {0747-5632},
  doi = {10.1016/j.chb.2021.106724},
  urldate = {2022-12-06},
  langid = {english}
}

@article{Hsieh-Sato2021,
  title = {Evaluation of Avatar and Voice Transform in Programming E-Learning Lectures},
  author = {Hsieh, Rex and Sato, Hisashi},
  year = {2021},
  month = jun,
  journal = {Journal on Multimodal User Interfaces},
  volume = {15},
  number = {2},
  pages = {121--129},
  issn = {1783-8738},
  doi = {10.1007/s12193-020-00349-5},
  urldate = {2022-06-27},
  langid = {english}
}

@inproceedings{Hu-etal2023,
  title = {Beyond {{Mirrors}}: {{Exploring Behavioral Changes}} through {{Comparative Avatar Design}} in {{VR Taiko Drumming}}},
  shorttitle = {Beyond {{Mirrors}}},
  booktitle = {Proceedings of the 29th {{ACM Symposium}} on {{Virtual Reality Software}} and {{Technology}}},
  author = {Hu, Yong-Hao and Hatada, Yuji and Narumi, Takuji},
  year = {2023},
  month = oct,
  series = {{{VRST}} '23},
  pages = {1--11},
  publisher = {Association for Computing Machinery},
  address = {New York, NY, USA},
  doi = {10.1145/3611659.3615690},
  urldate = {2024-11-04},
  isbn = {9798400703287}
}

@article{Isarida-Isarida2010,
  title = {Effects of Simple-and Complex-Place Contexts in the Multiple-Context Paradigm},
  author = {Isarida, Takeo and Isarida, Toshiko K},
  year = {2010},
  journal = {The Quarterly Journal of Experimental Psychology},
  volume = {63},
  number = {12},
  pages = {2399--2412},
  publisher = {SAGE Publications Sage UK: London, England},
  doi = {10.1080/17470211003736756}
}

@article{Kennedy-etal1993,
  title = {Simulator Sickness Questionnaire: {{An}} Enhanced Method for Quantifying Simulator Sickness},
  author = {Kennedy, Robert S and Lane, Norman E and Berbaum, Kevin S and Lilienthal, Michael G},
  year = {1993},
  journal = {The international journal of aviation psychology},
  volume = {3},
  number = {3},
  pages = {203--220},
  publisher = {Taylor \& Francis},
  doi = {10.1207/s15327108ijap0303_3}
}

@article{Koch-Coutanche2024,
  title = {Context Reinstatement Requires a Schema Relevant Virtual Environment to Benefit Object Recall},
  author = {Koch, Griffin E. and Coutanche, Marc N.},
  year = {2024},
  month = mar,
  journal = {Psychonomic Bulletin \& Review},
  issn = {1531-5320},
  doi = {10.3758/s13423-024-02472-w},
  urldate = {2024-03-13},
  langid = {english}
}

@inproceedings{Kruse-etal2023SUI,
  title = {The {{Influence}} of {{Virtual Agent Visibility}} in {{Virtual Reality Cognitive Training}}},
  booktitle = {Proceedings of the 2023 {{ACM Symposium}} on {{Spatial User Interaction}}},
  author = {Kruse, Lucie and Mostajeran, Fariba and Steinicke, Frank},
  year = {2023},
  month = oct,
  series = {{{SUI}} '23},
  pages = {1--9},
  publisher = {Association for Computing Machinery},
  address = {New York, NY, USA},
  doi = {10.1145/3607822.3614526},
  urldate = {2024-09-13},
  isbn = {9798400702815},
  postfix = {SUI}
}

@article{Lamers-Lanen2021,
  title = {Changing between Virtual Reality and Real-World Adversely Affects Memory Recall Accuracy},
  author = {Lamers, Maarten H. and Lanen, Maik},
  year = {2021},
  journal = {Frontiers in Virtual Reality},
  volume = {2},
  pages = {602087},
  issn = {2673-4192},
  doi = {10.3389/frvir.2021.602087}
}

@inproceedings{Latoschik-etal2017,
  title = {The Effect of Avatar Realism in Immersive Social Virtual Realities},
  booktitle = {Proceedings of the 23rd {{ACM Symposium}} on {{Virtual Reality Software}} and {{Technology}}},
  author = {Latoschik, Marc Erich and Roth, Daniel and Gall, Dominik and Achenbach, Jascha and Waltemate, Thomas and Botsch, Mario},
  year = {2017},
  month = nov,
  series = {{{VRST}} '17},
  pages = {1--10},
  publisher = {Association for Computing Machinery},
  address = {New York, NY, USA},
  doi = {10.1145/3139131.3139156},
  urldate = {2023-06-07},
  isbn = {978-1-4503-5548-3}
}

@article{Li-etal2016,
  title = {Social Robots and Virtual Agents as Lecturers for Video Instruction},
  author = {Li, Jamy and Kizilcec, Ren{\'e} and Bailenson, Jeremy and Ju, Wendy},
  year = {2016},
  month = feb,
  journal = {Computers in Human Behavior},
  volume = {55},
  pages = {1222--1230},
  issn = {0747-5632},
  doi = {10.1016/j.chb.2015.04.005},
  urldate = {2023-07-06},
  langid = {english}
}

@article{Lin-etal2023,
  title = {Measuring {{Interpersonal Trust}} towards {{Virtual Humans}} with a {{Virtual Maze Paradigm}}},
  author = {Lin, Jinghuai and Cronj{\'e}, Johrine and K{\"a}thner, Ivo and Pauli, Paul and Latoschik, Marc Erich},
  year = {2023},
  month = may,
  journal = {IEEE Transactions on Visualization and Computer Graphics},
  volume = {29},
  number = {5},
  pages = {2401--2411},
  issn = {1941-0506},
  doi = {10.1109/TVCG.2023.3247095}
}

@article{Makransky-etal2019JCAL,
  title = {A Gender Matching Effect in Learning with Pedagogical Agents in an Immersive Virtual Reality Science Simulation},
  author = {Makransky, Guido and Wismer, Philip and Mayer, Richard E.},
  year = {2019},
  journal = {Journal of Computer Assisted Learning},
  volume = {35},
  number = {3},
  pages = {349--358},
  issn = {1365-2729},
  doi = {10.1111/jcal.12335},
  urldate = {2024-12-18},
  copyright = {{\copyright} 2018 John Wiley \& Sons Ltd},
  langid = {english},
  postfix = {JCAL}
}

@inproceedings{Mizuho-etal2023ahs,
  title = {Virtual {{Omnibus Lecture}}: {{Investigating}} the {{Effects}} of {{Varying Lecturer Avatars}} as {{Environmental Context}} on {{Audience Memory}}},
  shorttitle = {Virtual {{Omnibus Lecture}}},
  booktitle = {Proceedings of the {{Augmented Humans International Conference}} 2023},
  author = {Mizuho, Takato and Amemiya, Tomohiro and Narumi, Takuji and Kuzuoka, Hideaki},
  year = {2023},
  month = mar,
  series = {{{AHs}} '23},
  pages = {55--65},
  publisher = {Association for Computing Machinery},
  address = {New York, NY, USA},
  doi = {10.1145/3582700.3582709},
  urldate = {2023-03-19},
  isbn = {978-1-4503-9984-5},
  postfix = {ahs}
}

@article{Mizuho-etal2023tvcg,
  title = {Effects of the {{Visual Fidelity}} of {{Virtual Environments}} on {{Presence}}, {{Context-dependent Forgetting}}, and {{Source-monitoring Error}}},
  author = {Mizuho, Takato and Narumi, Takuji and Kuzuoka, Hideaki},
  year = {2023},
  month = may,
  journal = {IEEE Transactions on Visualization and Computer Graphics},
  volume = {29},
  number = {5},
  pages = {2607--2614},
  issn = {1941-0506},
  doi = {10.1109/TVCG.2023.3247063},
  postfix = {tvcg}
}

@inproceedings{Mizuho-etal2023vrst,
  title = {Exploratory {{Study}} on the {{Reinstatement Effect Under}} 360-{{Degree Video-Based Virtual Environments}}},
  booktitle = {Proceedings of the 29th {{ACM Symposium}} on {{Virtual Reality Software}} and {{Technology}}},
  author = {Mizuho, Takato and Narumi, Takuji and Kuzuoka, Hideaki},
  year = {2023},
  month = oct,
  series = {{{VRST}} '23},
  pages = {1--2},
  publisher = {Association for Computing Machinery},
  address = {New York, NY, USA},
  doi = {10.1145/3611659.3617190},
  urldate = {2023-10-25},
  isbn = {9798400703287},
  postfix = {vrst}
}

@article{Mizuho-etal2024frontiers,
  title = {Multiple-Agent Promotion in a Grocery Store: Effects of Modality and Variability of Agents on Customer Memory},
  shorttitle = {Multiple-Agent Promotion in a Grocery Store},
  author = {Mizuho, Takato and Okafuji, Yuki and Baba, Jun and Narumi, Takuji},
  year = {2024},
  month = dec,
  journal = {Frontiers in Robotics and AI},
  volume = {11},
  publisher = {Frontiers},
  issn = {2296-9144},
  doi = {10.3389/frobt.2024.1397230},
  urldate = {2024-12-05},
  langid = {english},
  postfix = {frontiers}
}

@inproceedings{Mizuho-etal2024sap,
  title = {Investigating the {{Effects}} of {{Changing}} the {{Appearance}} of {{Screen-Based Avatars}} on {{Audience Memory}}},
  booktitle = {{{ACM Symposium}} on {{Applied Perception}} 2024},
  author = {Mizuho, Takato and Narumi, Takuji and Kuzuoka, Hideaki},
  year = {2024},
  month = aug,
  series = {{{SAP}} '24},
  pages = {1--9},
  publisher = {Association for Computing Machinery},
  address = {New York, NY, USA},
  doi = {10.1145/3675231.3675239},
  urldate = {2024-09-04},
  isbn = {9798400710612},
  postfix = {sap}
}

@article{Mizuho-etal2024tvcg1,
  title = {Reduction of {{Forgetting}} by {{Contextual Variation During Encoding Using}} 360-{{Degree Video-Based Immersive Virtual Environments}}},
  author = {Mizuho, Takato and Narumi, Takuji and Kuzuoka, Hideaki},
  year = {2024},
  journal = {IEEE Transactions on Visualization and Computer Graphics},
  pages = {1--14},
  issn = {1941-0506},
  doi = {10.1109/TVCG.2024.3403885},
  urldate = {2024-05-29},
  postfix = {tvcg1}
}

@article{Mizuho-etal2024tvcg2,
  title = {Multiple {{Self-Avatar Effect}}: {{Effects}} of {{Using Diverse Self-Avatars}} on {{Memory Acquisition}} and {{Retention}} of {{Sign-Language Gestures}}},
  shorttitle = {Multiple {{Self-Avatar Effect}}},
  author = {Mizuho, Takato and Takenaka, Shun and Narumi, Takuji and Kuzuoka, Hideaki},
  year = {2024},
  journal = {IEEE Transactions on Visualization and Computer Graphics},
  pages = {1--15},
  issn = {1941-0506},
  doi = {10.1109/TVCG.2024.3433498},
  urldate = {2024-09-04},
  postfix = {tvcg2}
}

@article{Oh-etal2018,
  title = {A {{Systematic Review}} of {{Social Presence}}: {{Definition}}, {{Antecedents}}, and {{Implications}}},
  shorttitle = {A {{Systematic Review}} of {{Social Presence}}},
  author = {Oh, Catherine S. and Bailenson, Jeremy N. and Welch, Gregory F.},
  year = {2018},
  month = oct,
  journal = {Frontiers in Robotics and AI},
  volume = {5},
  publisher = {Frontiers},
  issn = {2296-9144},
  doi = {10.3389/frobt.2018.00114},
  urldate = {2024-03-31},
  langid = {english}
}

@article{Patotskaya-etal2023,
  title = {Avoiding Virtual Humans in a Constrained Environment: {{Exploration}} of Novel Behavioural Measures},
  shorttitle = {Avoiding Virtual Humans in a Constrained Environment},
  author = {Patotskaya, Yuliya and Hoyet, Ludovic and Olivier, Anne-H{\'e}l{\`e}ne and Pettr{\'e}, Julien and Zibrek, Katja},
  year = {2023},
  month = feb,
  journal = {Computers \& Graphics},
  volume = {110},
  pages = {162--172},
  issn = {0097-8493},
  doi = {10.1016/j.cag.2023.01.001},
  urldate = {2024-06-22}
}

@inproceedings{Petersen-etal2021,
  title = {Pedagogical {{Agents}} in {{Educational VR}}: {{An}} in the {{Wild Study}}},
  shorttitle = {Pedagogical {{Agents}} in {{Educational VR}}},
  booktitle = {Proceedings of the 2021 {{CHI Conference}} on {{Human Factors}} in {{Computing Systems}}},
  author = {Petersen, Gustav B{\o}g and Mottelson, Aske and Makransky, Guido},
  year = {2021},
  month = may,
  series = {{{CHI}} '21},
  pages = {1--12},
  publisher = {Association for Computing Machinery},
  address = {New York, NY, USA},
  doi = {10.1145/3411764.3445760},
  urldate = {2024-12-17},
  isbn = {978-1-4503-8096-6}
}

@article{Pi-etal2022,
  title = {The Emotional Design of an Instructor: Body Gestures Do Not Boost the Effects of Facial Expressions in Video Lectures},
  shorttitle = {The Emotional Design of an Instructor},
  author = {Pi, Zhongling and Liu, Renjia and Ling, Hongjuan and Zhang, Xingyu and Wang, Shuo and Li, Xiying},
  year = {2022},
  month = jul,
  journal = {Interactive Learning Environments},
  volume = {0},
  number = {0},
  pages = {1--20},
  publisher = {Routledge},
  issn = {1049-4820},
  doi = {10.1080/10494820.2022.2105898},
  urldate = {2022-08-19}
}

@book{Reeves-Nass1996,
  title = {The Media Equation: {{How}} People Treat Computers, Television, and New Media like Real People},
  shorttitle = {The Media Equation},
  author = {Reeves, Byron and Nass, Clifford},
  year = {1996},
  publisher = {Cambridge University Press},
  address = {New York}
}

@inproceedings{Rzayev-etal2019,
  title = {The {{Effect}} of {{Presence}} and {{Appearance}} of {{Guides}} in {{Virtual Reality Exhibitions}}},
  booktitle = {Proceedings of {{Mensch}} Und {{Computer}} 2019},
  author = {Rzayev, Rufat and Karaman, G{\"u}rkan and Wolf, Katrin and Henze, Niels and Schwind, Valentin},
  year = {2019},
  month = sep,
  series = {{{MuC}} '19},
  pages = {11--20},
  publisher = {Association for Computing Machinery},
  address = {New York, NY, USA},
  doi = {10.1145/3340764.3340802},
  urldate = {2024-12-17},
  isbn = {978-1-4503-7198-8}
}

@inproceedings{Sanz-etal2015,
  title = {Virtual Proxemics: {{Locomotion}} in the Presence of Obstacles in Large Immersive Projection Environments},
  shorttitle = {Virtual Proxemics},
  booktitle = {2015 {{IEEE Virtual Reality}} ({{VR}})},
  author = {Sanz, Fernando Argelaguet and Olivier, Anne-H{\'e}l{\`e}ne and Bruder, Gerd and Pettr{\'e}, Julien and L{\'e}cuyer, Anatole},
  year = {2015},
  month = mar,
  pages = {75--80},
  issn = {2375-5334},
  doi = {10.1109/VR.2015.7223327},
  urldate = {2024-06-22}
}

@article{Schubert-etal2001,
  title = {The Experience of Presence: {{Factor}} Analytic Insights},
  author = {Schubert, Thomas and Friedmann, Frank and Regenbrecht, Holger},
  year = {2001},
  journal = {Presence: Teleoperators \& Virtual Environments},
  volume = {10},
  number = {3},
  pages = {266--281},
  publisher = {MIT Press},
  doi = {10.1162/105474601300343603}
}

@article{Shin-etal2021,
  title = {Context-Dependent Memory Effects in Two Immersive Virtual Reality Environments: {{On Mars}} and Underwater},
  author = {Shin, Yeon Soon and {Mas{\'i}s-Obando}, Rolando and Keshavarzian, Neggin and D{\'a}ve, Riya and Norman, Kenneth A},
  year = {2021},
  journal = {Psychonomic Bulletin \& Review},
  volume = {28},
  number = {2},
  pages = {574--582},
  publisher = {Springer},
  doi = {10.3758/s13423-020-01835-3}
}

@article{Smith-etal1978,
  title = {Environmental Context and Human Memory},
  author = {Smith, Steven M and Glenberg, Arthur and Bjork, Robert A},
  year = {1978},
  journal = {Memory \& Cognition},
  volume = {6},
  number = {4},
  pages = {342--353},
  publisher = {Springer},
  doi = {10.3758/BF03197465}
}

@article{Smith-Handy2014,
  title = {Effects of Varied and Constant Environmental Contexts on Acquisition and Retention},
  author = {Smith, Steven M and Handy, Justin D},
  year = {2014},
  journal = {Journal of Experimental Psychology: Learning, Memory, and Cognition},
  volume = {40},
  number = {6},
  pages = {1582--1593},
  publisher = {American Psychological Association},
  doi = {10.1037/xlm0000019}
}

@article{Smith-Handy2016,
  title = {The Crutch of Context-Dependency: {{Effects}} of Contextual Support and Constancy on Acquisition and Retention},
  author = {Smith, Steven M and Handy, Justin D},
  year = {2016},
  journal = {Memory},
  volume = {24},
  number = {8},
  pages = {1134--1141},
  publisher = {Taylor \& Francis},
  doi = {10.1080/09658211.2015.1071852}
}

@article{Smith-Vela2001,
  title = {Environmental Context-Dependent Memory: {{A}} Review and Meta-Analysis},
  author = {Smith, Steven M and Vela, Edward},
  year = {2001},
  month = jun,
  journal = {Psychonomic Bulletin \& Review},
  volume = {8},
  number = {2},
  pages = {203--220},
  publisher = {Springer},
  doi = {10.3758/BF03196157}
}

@article{Smith1984,
  title = {A Comparison of Two Techniques for Reducing Context-Dependent Forgetting},
  author = {Smith, Steven M},
  year = {1984},
  journal = {Memory \& Cognition},
  volume = {12},
  number = {5},
  pages = {477--482},
  publisher = {Springer},
  doi = {10.3758/BF03198309}
}

@incollection{Smith2013,
  title = {Effects of Environmental Context on Human Memory},
  booktitle = {The {{SAGE}} Handbook of Applied Memory},
  author = {Smith, Steven M},
  editor = {Perfect, T. J. and Lindsay, D. S.},
  year = {2013},
  pages = {162--182},
  publisher = {SAGE},
  address = {London}
}

@article{Smith2019,
  title = {Virtual Reality in Episodic Memory Research: {{A}} Review},
  author = {Smith, S Adam},
  year = {2019},
  journal = {Psychonomic Bulletin \& Review},
  volume = {26},
  number = {4},
  pages = {1213--1237},
  publisher = {Springer},
  doi = {10.3758/s13423-019-01605-w}
}

@inproceedings{Takenaka-etal2024,
  title = {Effects of {{Human}} and {{Animal Partner-Avatars}} on {{Profile Memory}} in {{Virtual Reality}}},
  booktitle = {{{ACM Symposium}} on {{Applied Perception}} 2024},
  author = {Takenaka, Shun and Mizuho, Takato and Narumi, Takuji and Kuzuoka, Hideaki},
  year = {2024},
  month = aug,
  series = {{{SAP}} '24},
  pages = {1--10},
  publisher = {Association for Computing Machinery},
  address = {New York, NY, USA},
  doi = {10.1145/3675231.3675241},
  urldate = {2024-09-04},
  isbn = {9798400710612}
}

@article{Tulving-Thomson1973,
  title = {Encoding Specificity and Retrieval Processes in Episodic Memory},
  author = {Tulving, Endel and Thomson, Donald M},
  year = {1973},
  journal = {Psychological Review},
  volume = {80},
  number = {5},
  pages = {352--373},
  publisher = {American Psychological Association},
  address = {US},
  issn = {1939-1471},
  doi = {10.1037/h0020071}
}

@article{Volonte-etal2016,
  title = {Effects of {{Virtual Human Appearance Fidelity}} on {{Emotion Contagion}} in {{Affective Inter-Personal Simulations}}},
  author = {Volonte, Matias and Babu, Sabarish V. and Chaturvedi, Himanshu and Newsome, Nathan and Ebrahimi, Elham and Roy, Tania and Daily, Shaundra B. and Fasolino, Tracy},
  year = {2016},
  month = apr,
  journal = {IEEE Transactions on Visualization and Computer Graphics},
  volume = {22},
  number = {4},
  pages = {1326--1335},
  issn = {1941-0506},
  doi = {10.1109/TVCG.2016.2518158},
  urldate = {2024-05-05}
}

@article{Walti-etal2019,
  title = {Reinstating Verbal Memories with Virtual Contexts: {{Myth}} or Reality?},
  author = {W{\"a}lti, Michel Juhani and Woolley, Daniel Graham and Wenderoth, Nicole},
  year = {2019},
  month = mar,
  journal = {PLOS ONE},
  volume = {14},
  number = {3},
  pages = {1--20},
  publisher = {Public Library of Science},
  doi = {10.1371/journal.pone.0214540}
}

@article{Weidner-etal2023,
  title = {A {{Systematic Review}} on the {{Visualization}} of {{Avatars}} and {{Agents}} in {{AR}} \& {{VR}} Displayed Using {{Head-Mounted Displays}}},
  author = {Weidner, Florian and Boettcher, Gerd and Arboleda, Stephanie Arevalo and Diao, Chenyao and Sinani, Luljeta and Kunert, Christian and Gerhardt, Christoph and Broll, Wolfgang and Raake, Alexander},
  year = {2023},
  month = may,
  journal = {IEEE Transactions on Visualization and Computer Graphics},
  volume = {29},
  number = {5},
  pages = {2596--2606},
  issn = {1941-0506},
  doi = {10.1109/TVCG.2023.3247072},
  urldate = {2024-12-10}
}

@inproceedings{Wobbrock-etal2011,
  title = {The Aligned Rank Transform for Nonparametric Factorial Analyses Using Only Anova Procedures},
  booktitle = {Proceedings of the {{SIGCHI}} Conference on Human Factors in Computing Systems},
  author = {Wobbrock, Jacob O. and Findlater, Leah and Gergle, Darren and Higgins, James J.},
  year = {2011},
  series = {{{CHI}} '11},
  pages = {143---146},
  publisher = {Association for Computing Machinery},
  address = {New York, NY, USA},
  doi = {10.1145/1978942.1978963},
  isbn = {978-1-4503-0228-9}
}

@article{Yuan-Gao2024,
  title = {Being {{There}}, and {{Being Together}}: {{Avatar Appearance}} and {{Peer Interaction}} in {{VR Classrooms}} for {{Video-Based Learning}}},
  shorttitle = {Being {{There}}, and {{Being Together}}},
  author = {Yuan, Quan and Gao, Qin},
  year = {2024},
  month = jul,
  journal = {International Journal of Human--Computer Interaction},
  volume = {40},
  number = {13},
  pages = {3313--3333},
  publisher = {Taylor \& Francis},
  issn = {1044-7318},
  doi = {10.1080/10447318.2023.2189818},
  urldate = {2024-09-14}
}

@inproceedings{Zibrek-etal2017,
  title = {Don't Stand so Close to Me: Investigating the Effect of Control on the Appeal of Virtual Humans Using Immersion and a Proximity-Based Behavioral Task},
  shorttitle = {Don't Stand so Close to Me},
  booktitle = {Proceedings of the {{ACM Symposium}} on {{Applied Perception}}},
  author = {Zibrek, Katja and Kokkinara, Elena and McDonnell, Rachel},
  year = {2017},
  month = sep,
  series = {{{SAP}} '17},
  pages = {1--11},
  publisher = {Association for Computing Machinery},
  address = {New York, NY, USA},
  doi = {10.1145/3119881.3119887},
  urldate = {2023-12-18},
  isbn = {978-1-4503-5148-5}
}

@article{Zibrek-etal2018,
  title = {The {{Effect}} of {{Realistic Appearance}} of {{Virtual Characters}} in {{Immersive Environments}} - {{Does}} the {{Character}}'s {{Personality Play}} a {{Role}}?},
  author = {Zibrek, Katja and Kokkinara, Elena and Mcdonnell, Rachel},
  year = {2018},
  month = apr,
  journal = {IEEE Transactions on Visualization and Computer Graphics},
  volume = {24},
  number = {4},
  pages = {1681--1690},
  issn = {1941-0506},
  doi = {10.1109/TVCG.2018.2794638},
  urldate = {2023-11-06}
}

@article{Zibrek-etal2019,
  title = {Is {{Photorealism Important}} for {{Perception}} of {{Expressive Virtual Humans}} in {{Virtual Reality}}?},
  author = {Zibrek, Katja and Martin, Sean and McDonnell, Rachel},
  year = {2019},
  month = sep,
  journal = {ACM Transactions on Applied Perception},
  volume = {16},
  number = {3},
  pages = {14:1--14:19},
  issn = {1544-3558},
  doi = {10.1145/3349609},
  urldate = {2023-11-06}
}

@article{Zibrek-etal2020,
  title = {The {{Effect}} of {{Gender}} and {{Attractiveness}} of {{Motion}} on {{Proximity}} in {{Virtual Reality}}},
  author = {Zibrek, Katja and Niay, Benjamin and Olivier, Anne-H{\'e}l{\`e}ne and Hoyet, Ludovic and Pettre, Julien and McDonnell, Rachel},
  year = {2020},
  month = nov,
  journal = {ACM Transactions on Applied Perception},
  volume = {17},
  number = {4},
  pages = {14:1--14:15},
  issn = {1544-3558},
  doi = {10.1145/3419985},
  urldate = {2024-05-05}
}

@inproceedings{Zibrek-McDonnell2019,
  title = {Social Presence and Place Illusion Are Affected by Photorealism in Embodied {{VR}}},
  booktitle = {Proceedings of the 12th {{ACM SIGGRAPH Conference}} on {{Motion}}, {{Interaction}} and {{Games}}},
  author = {Zibrek, Katja and McDonnell, Rachel},
  year = {2019},
  month = oct,
  series = {{{MIG}} '19},
  pages = {1--7},
  publisher = {Association for Computing Machinery},
  address = {New York, NY, USA},
  doi = {10.1145/3359566.3360064},
  urldate = {2023-11-05},
  isbn = {978-1-4503-6994-7}
}

\begin{IEEEbiography}[{\includegraphics[width=1in,height=1.25in,clip,keepaspectratio]{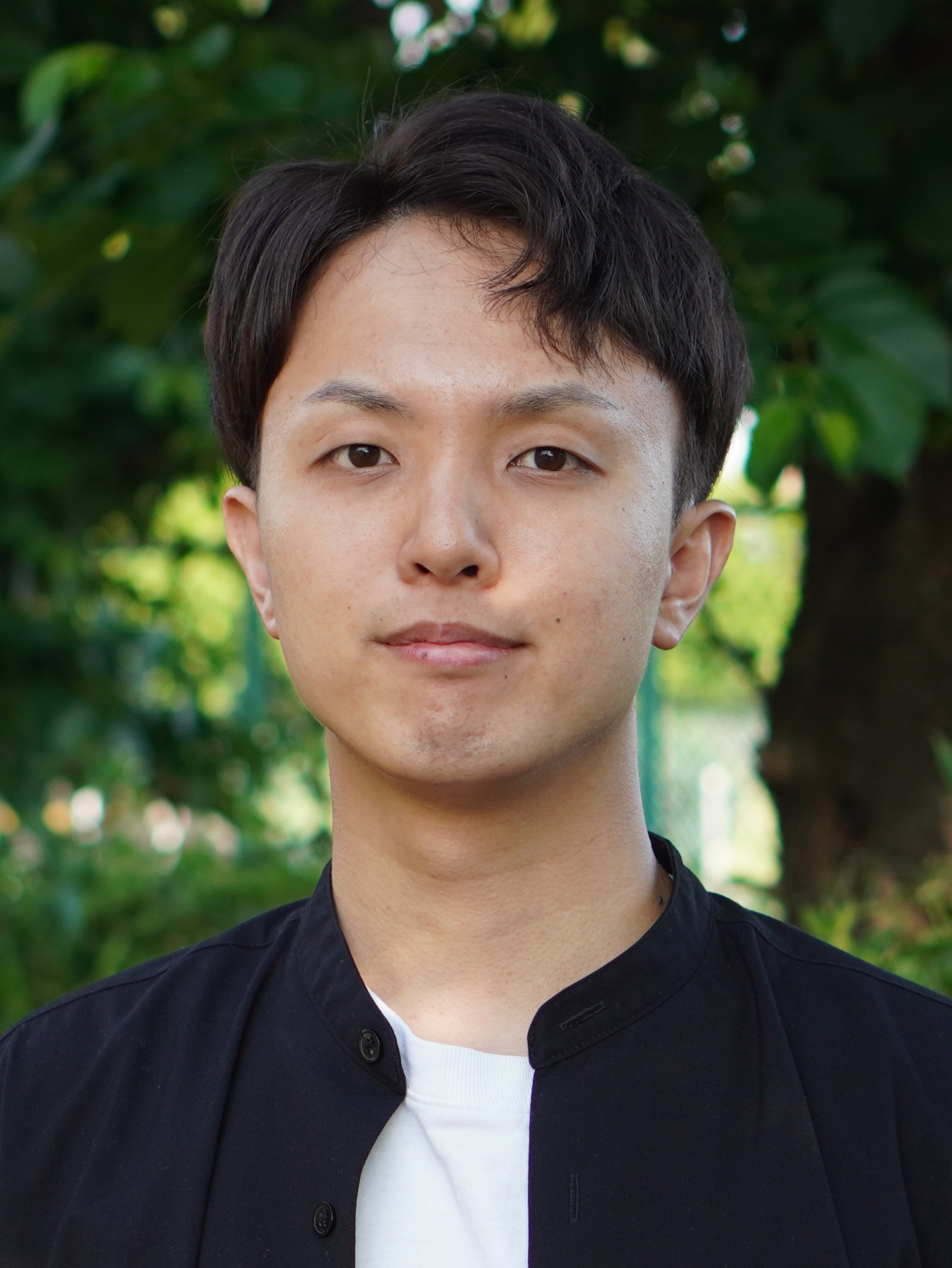}}]{Takato Mizuho}
is an assistant professor at the Graduate School of Information Science and Technology, the University of Tokyo. He received his B.S. and M.S. degrees from the University of Tokyo in 2020 and 2022, respectively.
He also received his Ph.D. in Information Science and Technology from the University of Tokyo in 2025.
His research interests include avatars, context-dependent memory, presence, and virtual reality.
\end{IEEEbiography}

\begin{IEEEbiography}[{\includegraphics[width=1in,height=1.25in,clip,keepaspectratio]{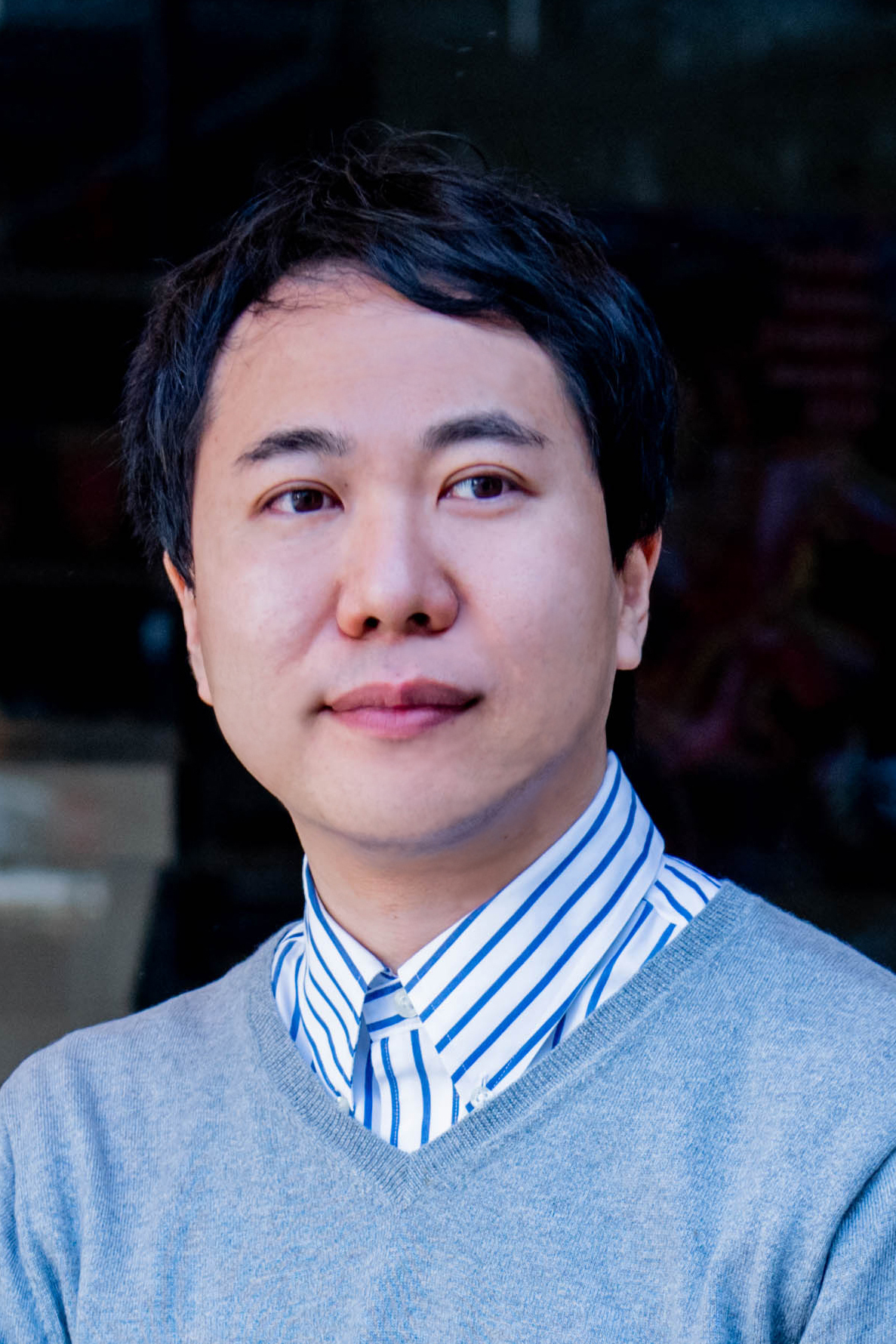}}]{Takuji Narumi}
is an associate professor at the Graduate School of Information Science and Technology, the University of Tokyo. He received his BE and ME degrees from the University of Tokyo in 2006 and 2008, respectively. He also received his Ph.D. in Engineering from the University of Tokyo in 2011. His research interests broadly include perceptual modification and human augmentation with virtual reality and augmented reality technologies.
\end{IEEEbiography}

\begin{IEEEbiography}[{\includegraphics[width=1in,height=1.25in,clip,keepaspectratio]{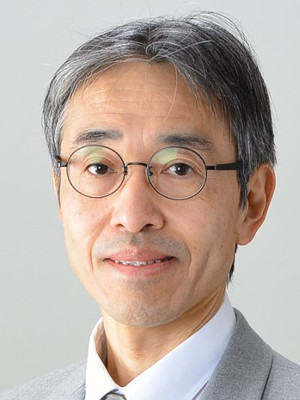}}]{Hideaki Kuzuoka}
is a professor at the Graduate School of Information Science and Technology, the University of Tokyo. He graduated from the Graduate School of Engineering, the University of Tokyo, in 1986 and received his Ph.D. in Engineering in 1992. His research interests include computer-supported cooperative work, social robotics, virtual reality, and human-computer interaction in general.
\end{IEEEbiography}

\vfill

\end{document}